\documentclass[12pt,preprint]{aastex}

\shorttitle{M81 Globular Cluster Photometry}
\shortauthors{Nantais et al.}

\begin{document}

\title{Hubble Space Telescope Photometry of Globular Clusters in  M81$^{\dagger}$}
\author{Julie B. Nantais}
\affil{Departamento de Astronom\'{i}a, Universidad de Concepci\'{o}n}
\affil{Av. Esteban Iturra s/n Barrio Universitario}
\affil{Casilla 160-C}
\affil{Concepcion, Chile}

\author{John P. Huchra, Andreas Zezas\altaffilmark{1,2}, Kosmas Gazeas, and Jay Strader}
\affil{Harvard-Smithsonian Center for Astrophysics}
\affil{60 Garden Street, Cambridge, MA 02138, USA}

\altaffiltext{1}{Dept. of Physics, University of Crete, Heraklion, Greece}
\altaffiltext{2}{Institute of Electronic Structure and LASER, Foundation for Research and Technology, Heraklion, Greece}

\footnotetext[\dagger]{Based on observations with the Hubble Space Telescope obtained at the Space Telescope Science Institute, operated by the Association of Universities for Research in Astronomy, Inc., under NASA contract NAS 5-26555.  These observations are associated with Program GO-10250 and Program GO-10584.} 


\begin{abstract}
We perform aperture photometry and profile fitting on 419 globular cluster (GC) candidates with $m_{V}$ $\leq$ 23 mag identified in {\it{Hubble Space Telescope}} Advanced Camera for Surveys {\it{BVI}} imaging, and estimate the effective radii of the clusters.  We identify 85 previously known spectroscopically-confirmed clusters, and newly identify 136 objects as good cluster candidates within the 3$\sigma$ color and size ranges defined by the spectroscopically confirmed clusters, yielding a total of 221 probable GCs.  The luminosity function peak for the 221 probable GCs with estimated total dereddening applied is $V\sim$(20.26 $\pm$ 0.13) mag, corresponding to a distance of $\sim$3.7$\pm$0.3 Mpc.  The blue and red GC candidates, and the metal-rich (MR) and metal-poor (MP) spectroscopically confirmed clusters, are similar in half-light radius, respectively.  Red confirmed clusters are about 6\% larger in median half-light radius than blue confirmed clusters, and red and blue good GC candidates are nearly identical in half-light radius.  The total population of confirmed and ``good'' candidates shows an increase in half-light radius as a function of galactocentric distance.
\end{abstract}

\section{Introduction}

The globular cluster (GC) system (GCS) of a galaxy provides valuable information on the nature of the major star-forming periods in the galaxy's history, including the initial collapse and any subsequent mergers.  The ages, locations, metallicities, and luminosity function of GCs can be used to determine the timing, formation methods, and destruction methods of the earliest star clusters, and to trace the most intense star formation periods in a galaxy's history.  Studies of GC populations in early-type galaxies have been used to understand how early-type galaxies form.

There are three classical GC formation scenarios applied to early-type galaxies: the gas-rich major merger scenario \citep{ash92}, the gas-poor minor merger scenario \citep{cot98}, and the in situ two-phase collapse scenario \citep{for97}.  All three models are intended to explain the common finding of two distinct metallicity (or color) subpopulations in the GCSs of massive galaxies: a red/metal-rich (MR) subpopulation and a blue/metal-poor (MP) subpopulation \citep{kun98,lee98,lar01a}.  Each classical scenario makes distinctive predictions about the MR and MP GC subpopulations in early-type galaxies.  In the major merger scenario, the MR GCs should be significantly younger than the MP GCs if the merger happened at a significantly later time than MP GC formation.  The MR GCs should be highly centrally concentrated while the MP GCs are not, resulting in an overall metallicity gradient in the GCS.  In the accretion scenario, both sets of GCs should be about the same age (depending on when the original host galaxies began to form), and only the mean metallicity of the MR GCs is expected to correlate with the luminosity of the host galaxy.  In the two-phase in situ formation scenario, the metallicities of both MR and MP clusters should relate to the luminosity of the host galaxy, and both MR and MP clusters should have size and metallicity gradients.  The blue in situ GCs would likely be just slightly older than the red in situ GCs.  A more modern scenario, the biased hierarchical merging scenario \citep{bea02,str05,rho05}, combines elements of all three classical scenarios.  The biased hierarchical merging scenario consists of a series of minor, gaseous mergers of gravitationally bound protogalaxies with two major epochs of GC formation.  The predictions for the MR and MP clusters are essentially the same as for the classical in situ scenario - blue GCs slightly older than red GCs but nearly coeval, and the metallicities of both correlating to that of the host galaxy.  However, biased hierarchical merging better reflects our current cosmological understanding of the early processes of galaxy formation.

The GCSs of spiral galaxies have not been as well-studied as those of elliptical galaxies, since their GCSs are generally poorer in total number and in number per unit galaxy magnitude (specific frequency).  Nonetheless, a number of studies have been performed to compare spiral galaxy GCSs to one another and to those of early-type galaxies.  Early-type spiral galaxies (S0/a-Sbc) such as the Milky Way \citep{zin85} and Andromeda \citep{bar00} have been found to have bimodal GC metallicity distributions.   Late-type spirals (Sbc-Sm), such as the Sc galaxy M51 \citep{cha04}, sometimes show evidence for only a MP GC population.  As with early type galaxies, the mean GC metallicity correlates with the luminosity of the host galaxy \citep{bro91}.

The GC population of M81 was first studied photometrically in depth by \citet[PR95]{pr95}, who performed a $BVR$ ground-based optical survey of cluster candidates in M81.  PR95 found about 70 unresolved GC candidates and 20 resolved GC candidates with $V\leq$ 21 mag.  They found a GC luminosity function (GCLF) similar to those of the Milky Way and M31, with a turnover magnitude at $M_{V0}$ = 20.3 mag, corresponding to an absolute magnitude of $-7.37$ using the \citet{fre01} distance modulus (m-M = 27.80 mag).  The next major optical survey of M81 star cluster candidates was performed by \citet[CFT01]{cft01} using {\it{Hubble Space Telescope}} {\it{(HST)}} Wide Field Planetary Camera 2 (WFPC2) $BVI$ imaging.  Their survey covered 40 square arcminutes.  They found 114 compact star cluster candidates, about 59 of which appeared to be old GCs upon color analysis in \citet{ctf01}.  A third survey of the M81 GCS was performed by \citet{cha04}, again using WFPC2.  Their images were primarily in the $B$, $V$, and $I$ filters, with $U$ and $R$ imaging available in some fields.  They identified 47 clusters with a mean color similar to that of red GCs in early-type galaxies.  The M81 GCLF had a turnover magnitude of M$_V$ $\sim$ -7.5 in the \citet{cha04} study.

In \citet{nan10a}, we created a catalog of star cluster candidates using F814W ($I$-band) images from the {\it{HST}} Advanced Camera for Surveys (ACS).  We identified over 2500 objects including H {\small{II}} regions, galaxies, GC candidates, and ambiguous objects, and performed a rudimentary color analysis on some of the objects using MMT Megacam $g$- and $r$-band imaging.  In \citet{nan10b}, we obtained spectra of 74 GCs, including 62 that had never been previously confirmed, and analyzed the metallicities and kinematics of the 108 spectroscopically confirmed M81 GCs.  In this paper, we continue our analysis of the M81 GCS with {\it{HST}} ACS $BVI$ photometry and radial profile fitting of 419 GC candidates.  In Section 2, we describe our data and explain the initial selection of GC candidates.  In Section 3, we discuss the completeness limits of our data.  In Section 4, we describe our methods of obtaining $BVI$ photometry and of classifying GC candidates according to color and size.  In Section 5 we describe the results, including colors, GCLF, and the half-light radii of MP and MR clusters.  Finally, in Section 6 we provide a summary of our analysis.

\section{Data and Initial Source Selection}
We use {\it{HST}} ACS Wide Field Camera (WFC) imaging of M81 in the F435W ($B$), F606W ($V$), and F814W ($I$) passbands.  The $B$ and $V$ imaging comes from {\it{HST}} program GO-10584 (P.I. Andreas Zezas).  The $I$-band imaging comes primarily from {\it{HST}} program GO-10250 (P.I. John Huchra) with five additional images from GO-10584.  GO-10584 has 29 fields in $B$ and $V$ and five fields in $I$, and GO-10250 (the dataset used in \citet{nan10a}) has 24 fields in I.  The total range of the data almost entirely covers the visible major and minor axes of M81, as defined in \citet{rc3}.  Figure 1 shows the region covered by the {\it{HST}} ACS mosaics, with $B$ and $V$ coverage represented by a solid line and $I$-band coverage by a dotted line.  At the distance of M81, 1 arcmin = 0.99 kpc.  Table 1 provides basic exposure info for all fields in $B$, $V$, and $I$: the {\it{HST}} proposal number, field ID, central right ascension and declination for each field, filter of observation (F435W = ``$B$'', F606W = ``$V$'', F814W = ``$I$''), and total exposure time of the field.  A full version of Table 1 will be available online.

The images in all three bands were processed using standard STScI pipeline processing (CALACS $+$ Multidrizzle), but were not background-subtracted.  The non-background-subtracted counts-per-second images, and the corresponding {\it{HST}} weight images produced by Multidrizzle, were combined using SWARP \citep{ber08}.  The X/Y and right ascension/declination coordinates were carefully matched among the individual pointings in each mosaic image and among the images in each band.  Since the total M81 mosaic is very large, the final image was separated into a northern mosaic and a southern mosaic, with a small area of overlap between the two at approximately the declination of the M81 nucleus.  The final astrometry of the V-band images, from which we derive our coordinates, differed from a sample of 132 Sloan Digital Sky Survey (SDSS) stars by a mean of -0.47$\arcsec$ in right ascension and -0.36$\arcsec$ in declination, with a dispersion of 0.11$\arcsec$ in right ascension and 0.13$\arcsec$ in declination.  A similar astrometric comparison was made with 111 2MASS All-Sky Point Source Catalog stars.  The {\it{HST}} coordinates were offset from the 2MASS coordinates by $+0.3\arcsec$ with a dispersion of $0.2\arcsec$ in both right ascension and declination, with a total mean offset of $+0.5\arcsec$.

Initial catalogs of sources were created using Source Extractor \citep{ber96} on the $B$, $V$, and $I$ mosaics separately, and sources in all three catalogs were later matched.  The detection and analysis thresholds in $V$ and $I$ were chosen to be 15 $\sigma$ above the local background, so as to screen out local maxima, substructures of galaxies and H {\small{II}} regions, and spurious detections.  The screening of spurious detections is especially important near the M81 nucleus.  Since we expect our GC distribution to be centrally concentrated, greater detection efficiency in the inner regions granted by a 15 $\sigma$ limit is preferable to better detection of faint sources in the outer regions (most of which will probably be galaxies) that would be granted by a lower detection threshold such as 5 $\sigma$.  In the $B$-band, a 5 $\sigma$ threshold for detection and analysis was chosen, since the $B$-band images have much lower diffuse background and therefore less need for spurious detection screening than the $V$- and $I$-band images.  The properties returned by Source Extractor for individual objects include the elongation, coordinates (X/Y and right ascension/declination), {\it{HST}} VEGAMAG magnitudes (using flexible Kron radii), aperture magnitudes estimated with a 10 pixel aperture radius (used to determine colors when objects in the catalogs were matched), and CLASS\_STAR values to indicate point-like or extended shape.  The 15 $\sigma$ limits on the $V$- and $I$-band catalogs allowed for point sources (CLASS\_STAR $>$ 0.8) to be detected to $\sim$24.7 and 24.2 mag, respectively.  The 5$\sigma$ limit on the $B$-band catalog allowed for detection of point sources down to $\sim$25.7 mag.

By identifying all extended, spectroscopically confirmed GCs from \citet{nan10b}, \citet{sch02}, \citet{pbh95}, and \citet{bro91} in our $BVI$ images and Source Extractor catalogs, we built a working sample of 86 clusters to determine appropriate shape, color, and size cuts to identify GC candidates. (One spectroscopically confirmed cluster, Object 2219 from \citet{nan10a}, was not found in the $B$ and $V$ images.  It lies within one of the several triangular gaps in the $B$ and $V$ coverage.)  Although there are 108 objects identified as spectroscopically confirmed GCs in M81, 8 failed to make the size cut in \citet{nan10a} (and thus using them might lead to misidentifying many stars and asterisms as globular clusters), and the remaining 14 are outside the spatial coverage of the images.  For shape and size criteria, we settled on elongation $\leq$ 2 and CLASS\_STAR $\leq$ 0.05.  The elongation limit excludes very non-round objects like some elliptical galaxies and cosmic rays.  The CLASS\_STAR limit excludes very round but very concentrated objects like the eight starlike spectroscopic ``clusters'' in \citet{nan10a}.  (These 8 clusters are not used in creating our color range.)  All confirmed clusters besides those eight objects had CLASS\_STAR $\leq$ 0.05.  Most confirmed clusters have CLASS\_STAR of about 0.03, regardless of their magnitude (and have magnitudes down to $V$ of about 21).  Among other candidates, fainter objects ($V$ $>$ 21) are more likely to have CLASS\_STAR of 0 or 0.01 than are brighter objects.

While matching objects between Source Extractor catalogs in $B$, $V$, and $I$, we allowed an initial maximum $V-I$ (VEGAMAG) of 2.0.  Then we limited our candidates to objects detected in all three images and with $V$ $\leq$ 23, about 2.8 mag past the estimated GC luminosity function (GCLF) peak of $V$ = 20.2.  This helps reduce contamination by faint background galaxies (which may be $B$-band dropouts) and spurious detections while retaining $\gtrsim$ 95\% of the GCs if $\sigma_{GCLF} \sim 1.2$ mag.  Finally, we used the confirmed GCs to determine a color cut.  We determined the average 10-pixel-radius colors (using circular radii in Source Extractor) and ($B-V$)--($V-I$) color ratio of the clusters (to constrain their position on a color-color diagram to within a certain range typical of GCs) and excluded all objects outside at least one of the three 5 $\sigma$ color ranges.  The \citet{bro91} spectroscopic cluster HS-02 (Object 2304 in our \citet{nan10a} $I$-band catalog) was so blue as to be outside the 5 $\sigma$ color ranges determined by the full set of 86 clusters.  Upon visual inspection we noticed that, while it is a genuine star cluster of some kind, it has a somewhat irregular appearance compared to other GCs.  Figure 2 shows BH91 HS-02 along with four other spectroscopically confirmed GCs.  Between this irregular appearance and its very blue color, we suspect that BH91 HS-02 is a young, luminous cluster misidentified as a GC in the low-resolution spectroscopy of BH91.  We exclude this cluster from further analysis and ultimately from our $BVI$ catalog.

After excluding BH91 HS-02 from our reference list of confirmed GCs, we refined our 5 $\sigma$ color range for GC selection using the remaining 85 clusters.  We retained objects with Source Extractor VEGAMAG colors of 0.36 $\leq$ ($B-V$) $\leq$ 2.11, 0.36 $\leq$ ($V-I$) $\leq$ 1.55, and -0.13 $\leq$ ($B-V$)--($V-I$) $\leq$ 0.69.  Finally, after performing the color cuts, we visually inspected all objects, eliminating the most obvious open clusters, H {\small{II}} regions, blends, galaxies, spurious detections, and some tiny point-like sources.  Figure 3 shows eight examples of rejected objects, including a galaxy, an H {\small{II}} region, and several faint blends and fuzzy objects unlikely to be GCs.  The remaining objects, 419 in all, comprise our final catalog of cluster candidates.  The coordinates of these objects, along with their matches in CFT, PR95, and \citet{nan10a}, and their spectroscopic confirmation status from \citet{pbh95}, \citet{sch02}, \citet{bro91}, and \citet{nan10b}, are shown in Table 2.  (A full version of Table 2 will be available online.)

Before any analysis, there were 255,907 candidate objects detected in the V-band images by Source Extractor.  Of these, about 4\% (about 11,800 objects) were left over after making CLASS\_STAR, magnitude, and elongation cuts.  Detecting all objects in all 3 images within a 10 pixel matching radius narrowed this number to about 3750 objects, or 1.5\% of the original total object number.  Making the 3$\sigma$ color and color-color-relation cuts based on the confirmed clusters as described above, about 1960 objects, or 0.77\% of the original candidates, are left.  The remaning objects are eliminated by visual inspection and stricter matching radius limits (on the order of 1.5-2 pixels) based on the typical range of right ascension and declination offsets of well-matched objects, and the final 419 objects represent 0.16\% of the original Source Extractor catalog objects.

In \citet{nan10b}, we spectroscopically confirmed over 40 H {\small{II}} regions and a few background galaxies and open clusters, as well as over 60 previously unconfirmed GCs.  Of the non-GC spectra from \citet{nan10b}, seven objects have been identified among our 419 GC candidates: four H {\small{II}} regions and three background galaxies.  These confirmed non-GCs are shown in Figure 4.  All of the H {\small{II}} regions except for Object 57 look like typical GCs.  The galaxies are somewhat more irregular in appearance but are also not obvious non-clusters.  The presence of only 7 confirmed non-GCs compared to 85 confirmed GCs among spectroscopically identified candidates suggests an $\sim$8\% contamination rate among bright ($V\gtrsim$ 21) GC candidates.

\section{Completeness}
To determine the completeness levels in our catalog, we followed a method similar to that used in \citet{nan10a}.  We took two 6000 $\times$ 12000 pixel subsets from the northern and southern halves of the $B$, $V$, and $I$ mosaics reaching out to 10$\arcmin$ from the nucleus of M81.  We overlaid a repeating grid of images of 10 bright M81 GCs, typical in FWHN and elongation among the spectroscopically confirmed M81 GC sample, onto these mosaic subsections, dimmed to varying magnitudes using IMARITH in IRAF$^{1}$.  We first ran Source Extractor on the subsections without the overlaid clusters (effectively recreating the original catalog for these smaller images), to determine the locations of objects that could be confused with the artificial clusters during the retrieval process.  Then we ran Source Extractor on the images with overlaid clusters, and looked for matches within 5 pixel to all overlaid clusters that were not themselves within 5 pixel of an object on the original image.  To minimize the inflation of completeness levels by new spurious detections created in the overlay process, we made sure to count only one coordinate match per artificial cluster position.  We performed the same CLASS\_STAR and ellipticity cuts on our Milky Way models as on our real globular clusters, although since we used circular models for the fake clusters, the ellipticity would not have been a significant constraint.

\footnotetext[1]{IRAF is distributed by the National Optical Astronomy Observatory, which is operated by the Association of Universities for Research in Astronomy, Inc., under cooperative agreement with the National Science Foundation.}

Figure 5 shows the completeness in $V$ (VEGAMAG) as a function of magnitude for different distances from the center of M81.  At our cutoff of $V$ = 23 mag, the average completeness is $\sim$ 71\%, and lowest in the inner regions.  Figure 6 shows the completeness limits for the $B$ and $I$ bands.  For $I$-band, completeness drops to 80\% (lower toward the inner regions) at $I$ $\sim$ 22 mag, while in the $B$-band, completeness fades to 80\% shortly after $B$ = 24.5 mag.

In \citet{nan10a}, we constructed \citet{kin66} models of high-latitude Milky Way GCs using the structural data in \citet{har96}$^{2}$ to gauge the effects of size cuts (based on a minimum Gaussian FWHM) on cluster retention.  We determined that no more than about 10\% of M81 globulars were likely to be rejected due to our FWHM limit.  Since we used CLASS\_STAR (which is a Source Extractor parameter derived from comparing the profile of the object to an estimated stellar point spread function) in place of a FWHM cut in this study, we tested the retention of Milky Way GC King models in the southern subsection subject to a CLASS\_STAR $\leq$ 0.05 cut at 18th magnitude.  At very bright magnitudes, cluster retention is dependent on the Gaussian FWHM, but not surface brightness.  Again, we find that retention of the Milky Way King models is only reduced by about 10\% when we apply a CLASS\_STAR = 0.05 upper limit.  Therefore, fewer than 10\% of GCs are unable to be detected due to having too high a CLASS\_STAR, assuming a distribution of morphologies similar to that of the Milky Way GCs.  These results are similar to if we use the Gaussian FWHM itself to make the cut.

\footnotetext[2]{We use the December 2010 update of the catalog, located at http://physwww.mcmaster.ca/$\sim$harris/mwgc.dat}

We also found in \citet{nan10a} that another $\sim$ 10\% of GCs may be lost due to low surface brightness at intermediate and faint magnitudes.  We therefore test the retention of the Milky Way King models in the southern subsection at $V$ = 22 mag, one magnitude above our $V$-band cutoff.  Detection rates at $V$ = 22 mag, not considering surface brightness, are still near 100\%.  We apply the CLASS\_STAR $\leq$ 0.05 cut here as well.  This time we find up to 45\% additional loss due to low surface brightness.  After both CLASS\_STAR and surface brightness losses, only 45\% of Milky Way-based King models are retained.  This is due to the stronger surface brightness limits imposed by a 15 $\sigma$ detection limit as opposed to a 5$\sigma$ detection limit.  With a 5 $\sigma$ detection limit, there would be 20\% additional loss when a CLASS\_STAR cut is applied, leading to a $\sim$ 70\% cluster retention at $V$ = 22 mag, similar to \citet{nan10a}.  However, we would not expect the lowest-surface-brightness GCs to pass visual inspection as GCs.  Some low-surface-brightness GCs may resemble open clusters or just barely noticeable concentrations of stars.  We will continue to use the 15 $\sigma$ detection limit, but have performed this additional completeness analysis to account for the surface brightness losses predicted by the Milky Way King models.  Figure 7 shows the Milky Way King model completeness limits as a function of magnitude in the $V$-band, fading gradually with increasing magnitude.  The more gradual fading in the inner regions in Figures 5-7 is most likely due to a high rate of spurious detections (chance superpositions of stars, etc.) typical of the crowded bulge areas with high local background variability.

\section{Photometry and Cluster Classification}
We used the APPHOT task in IRAF to perform aperture photometry on our 419 GC candidates.  We used the $V$-band images to obtain integrated magnitudes, and chose integrated magnitude aperture radii based on where the radial profile of the cluster light blended into the background.  We used smaller radii for objects with low signal-to-noise and which blended in to the local background rapidly, and larger radii for objects that blended in to the local background more gradually.  Since we expect the GCs to have varying profiles, we did not perform aperture corrections.   We also tailored the aperture radii to the environment near each cluster, sometimes using a smaller aperture to minimize noise or light from neighboring objects.  A radius of 20 pixels (1$\arcsec$) was sufficient for most clusters, although we used radii between 10 and 60 pixels depending on the size and environment of the object.  Local background levels were determined in an annulus beginning at 40 pixels (2$\arcsec$) from the center of the object, with a width of 20 pixels.  For objects with photometric aperture radii larger than 40 pixels, the inner radius of the background annulus was equal to the photometric aperture radius.  The width of the background annulus was again 20 pixels.

We calculated growth curves for 20 typical-color and typical-shape GCs out to a radius of 20 pixels, enough to get a reasonable estimate of the total magnitude of most GCs.  The mean absolute difference between the colors at 10 pixel and those at 20 pixels was only 0.02 mag, with a dispersion of 0.03 mag for $B-V$ and 0.02 mag for $V-I$.  The 10-pixel colors of confirmed GCs showed a tight color-color relation.  Thus, we used 10 pixel (0.5$\arcsec$) color radii for all objects, for the sake of simplicity and consistency.  The 10-pixel color radius resulted in a low aperture correction for most GCs, yet provided precise and accurate enough colors for low-scatter color-color relations.

While the point spread functions (PSFs) in the ACS $B$ and $V$ images were similar, the PSF in the $I$-band was slightly broader, creating a potentially significant systematic offset in $V-I$ and $B-I$ colors.  We convolved simple King models based on Milky Way clusters with the PSFs in each band, and estimated colors for the model clusters with 10-pixel aperture radii in each band.  The $B-V$ color was only about 0.01 mag bluer than expected, but the $V-I$ color was about 0.04 mag bluer.  To correct for the effect of PSFs on color, we convolved our $B$-band and $V$-band GC candidate images with a kernel to match them to the $I$-band PSF.  First, we determined empirical PSFs for the $B$-, $V$-, and $I$-band images from 62 bright, isolated, non-saturated point sources (the same point sources in each band) with the IRAF task PSF in the DAOPHOT package.  Next, we used the IRAF task PSFMATCH in the IMAGES.IMMATCH package to create $B$-to-$I$ and $V$-to-$I$ convolution kernels.  Finally, we used the IRAF task FCONVOLVE in the STSDAS.ANALYSIS.FOURIER package to alter 300 $\times$ 300 pixel $B$ and $V$ thumbnail images of the GC candidates to match the $I$-band PSF.  We then performed 10-pixel radius aperture photometry on the altered $B$ and $V$ thumbnail images and the unaltered $I$-band images to determine colors.  Total $V$ magnitudes were determined from the unaltered $V$ images.  Performing these alterations on the artificial Milky Way King model images reduced the systematic $V-I$ color offset by about 0.03 mag.

We determined uncertainties in magnitudes and colors by scaling the source count values in the mosaic images and sky background dispersions according to the mean effective exposure times (determined from the weight images) in the apertures and annuli.  With the relevant quantities properly scaled, we used IRAF's equation for photometric uncertainty:

\begin{equation}
Error = (flux/gain + area \times stdev^2 + area^2 \times stdev^2/nsky)^{1/2}
\end{equation}

In the above equation, $nsky$ is the number of pixels in the sky annulus, $area$ is the number of pixels in the photometric aperture, $gain$ is the gain of the CCD in electrons, $flux$ is the flux of the object in counts, and $stdev$ is the dispersion in the sky background in counts.  As in \citet{nan10a}, we assumed a minimum uncertainty of 0.02 mag, and added this minimum uncertainty in quadrature to all uncertainties.  Also, we flagged all objects in which either the photometric aperture or the sky annulus spanned regions of widely differing effective exposure time.

To transform our {\it{HST}} magnitudes to the Johnson-Cousins scale, we used the iterative synthetic transformations from Table 32 of \citet{sir05}.  $B$ and $V$ magnitudes were determined using the $B-V$ color and the $I$ magnitude was determined using the $V-I$ color.

Once we determined the colors of all 419 cluster candidates, we used the spectroscopically confirmed clusters to define 3 $\sigma$ color ranges for classification of the best GC candidates.  The 3 $\sigma$ color ranges were 0.58 $\leq$ $B-V$ $\leq$ 1.37 and 0.77 $\leq$ $V-I$ $\leq$ 1.69, and the 3-sigma color-color relation range was $-0.39$ $\leq (B-V)-(V-I) \leq -0.12$.  Two hundred seventy-nine of our 419 objects (including all 85 confirmed clusters) fit within all of these color ranges.  The remaining 140 objects outside any one of these color ranges were retained as poor candidates.  The nature of these candidates is discussed in Section 5.1.

As \citet{cft01} note, background spiral galaxies often have similar colors to GCs.  Of 57 visually-identified galaxies rejected from our cluster catalog on the basis of morphology, 38 had colors and positions on the color-color diagram within the 3 $\sigma$ $B-V$, $V-I$, and ($B-V$)--($V-I$) ranges defined by the spectroscopically confirmed clusters.  However, background galaxies vary more widely in apparent size than GCs, and easily reach large apparent sizes.  To reduce pollution from background galaxies in our best GC candidate sample, we implemented a 3 $\sigma$ size cut based on the FWHMs of confirmed GCs computed with IRAF's RADPROF task.  We found the mean FWHM of a confirmed GC to be 3.85 pixel, with a $\sigma$ of 2.65 pixel.  With these results, we defined a 3 $\sigma$ FWHM upper limit of 11.8 pixel.  All objects within our 3 $\sigma$ color and size ranges were labeled as ``good'' candidates, while all objects within our 3 $\sigma$ color ranges but outside the 3 $\sigma$ size range were labeled as ``fair'' candidates.  All but two of the 85 confirmed GCs fall within the 3 $\sigma$ upper size limit, and 4 of the 7 confirmed non-GCs also fall within the size limit.  For purposes of further analysis such as GCLF calculation, ``fair'' candidates are assumed to be probable background galaxies, and ``poor'' candidates are assumed to be either galaxies or other non-GC objects such as young clusters and H {\small{II}} regions.  ``Conf'' candidates are the 85 spectroscopically confirmed GCs, and ``nongc'' are objects similar to GC candidates in appearance and color which have been spectroscopically confirmed as either H {\small{II}} regions or galaxies in \citet{nan10b}.  Table 3 (full version will be available online) gives the photometry of the objects, uncorrected for reddening.  The quality flags defined above are listed for each object in Table 3.  Also included are flags denoting whether the candidate fell on an edge or intersection between images within the photometric aperture or sky annulus.

Figure 8 shows five examples each of ``good'' (top row) ``fair'' (middle row) and ``poor'' (bottom row) GC candidates, arranged in order of increasing magnitude (brightest to faintest).  The brighter ``good'' objects resemble typical spectroscopically confirmed GCs, while the fainter ones appear more ambiguous and possibly open-cluster-like.  The ``fair'' candidates have appearances consistent with diffuse galaxies or possibly star clusters, although they are not analyzed as clusters.  The ``poor'' candidates tend to have more amorphous appearances consistent with galaxies or young objects (H {\small{II}} regions, OB associations, and open clusters).

\section{Results}
\subsection{Radial Distribution of GC candidates}
To understand whether the spatial distributions of our ``good,'' ``fair,'' and ``poor'' candidates differ significantly, we estimated the density of each type of object per square arcminute as a function of projected distance from the center of M81.  First, we estimated the total projected area in arcminutes within concentric circular annuli of 1$\arcmin$ width centered on the nucleus of M81 and bound by the corners of the $I$-band mosaic.  We then counted the number of each type of cluster candidate (``conf,'' ``good,'' ``fair,'' and ``poor'') within each annulus.  Finally, we divided the number of candidates by the area of the annulus in square arcminutes, and multiplied by a fractional completeness estimate for each distance bin to correct for how completeness varies as a function of distance.  The fractional completeness factor was estimated by averaging the mean fractional completeness found in each distance bin at $V$ = 18, 21, and 23 mag (see Figure 5), since the total sample contains bright, intermediate-magnitude, and faint objects.  The innermost 1$\arcsec$ had acompleteness factor of about 83\%, and the other distance bins had completeness factors of about 90-91\%.

Figure 9 shows the density of different types of objects as a function of projected distance from the M81 nucleus.  The ``good'' candidates (solid line), those that meet our color and size criteria based on the confirmed candidates, show a centrally concentrated distribution that fades gradually at projected distances  $>5\arcmin$.  This distribution is consistent with most objects at small radii being genuine GCs, and most objects at large radii being a mixture of GCs and background galaxies.  The spectroscopically confirmed clusters (long-dashed line) show strong central concentration that is ``cut off'' at projected distances less than $\sim$2$\arcmin$.  This shortage of confirmed clusters near the nucleus is likely a selection bias due to the difficulty of obtaining GC spectra near the nucleus.   In the \citet{nan10b} study, our ability to obtain GC spectra was spatially limited by the finite fiber separation and individual fiber widths of the Hectospec instrument.  Few confirmed clusters are found at large projected radii.  ``Fair'' cluster candidates (dotted line) have a nearly flat distribution, with the largest concentrations at small and intermediate projected radii, and little drop-off at large projected radii.  Some of the inner ``fair'' candidates may be star clusters, but many ``fair'' candidates at greater projected distances may be galaxies.  Finally, the ``poor'' candidates (short-dashed line) show the largest concentrations at intermediate projected distances, indicating that many of these objects may be young massive clusters and compact H {\small{II}} regions.

\subsection{Colors and Reddening}
Figure 10 shows a color-color plot of our good and spectroscopically confirmed GC candidates along with the colors of Milky Way GCs, and linear fits to the Milky Way clusters, spectroscopically confirmed M81 GCs, and good but unconfirmed M81 GC candidates.  Our $V-I$ colors (Y-axis) are systematically bluer than Milky Way $V-I$ colors by about 0.03 mag, despite the fact that we corrected for the PSF differences between the $V$-band and the $I$-band.  The mean $V-I$ of our spectroscopically confirmed M81 GCs is 1.23 mag, while the mean $V-I$ of Milky Way clusters with $E(B-V)$ $<$ 0.5 is 1.27 mag.  We investigated several possible ways to eliminate the remaining color offset, including larger photometric apertures, larger and more distant sky annuli, charge transfer efficiency corrections, removal of foreground reddening from system magnitudes, and adjustments to the \citet{sir05} zeropoints equal to the changes in the VEGAMAG zeropoints since the Sirianni et al. publication.  None of these adjustments to our photometry eliminated or significantly reduced the remaining slight offset from Milky Way $V-I$ colors in our spectroscopically confirmed GCs.  It is possible that the remaining color offset is related to the differences in wavelength range, and thus the extinction (which is heavily dependent on wavelength), between Johnson-Cousins and {\it{HST}} filters.  However, correcting for foreground reddening in the system magnitudes prior to transformation did not notably alleviate the color offset, and precise total reddening estimates for individual clusters would be difficult to obtain.  

As a simple test on our colors, we identified 127 of our globular cluster candidates, including 42 ``good'' candidates and 36 confirmed clusters, in the Sloan Digital Sky Survey (SDSS) Data Release 7.  We used the \citet{lup05} transformations, specifically those using the ``redder'' SDSS magnitudes and colors ($g$ and $r$ to determine $B$, $r$ and $i$ to determine $V$, and $i$ and $z$ to determine $I$) to convert SDSS colors into Johnson $BVI$ colors.  The SDSS objects had similar colors in $B-V$ and $V-I$ to what we calculated, with a mean offset of $\sim -$0.02 mag in $B-V$ and $\sim -$0.01 mag in $V-I$ for confirmed and ``good'' candidates and $\sim -$0.02 mag in both colors for all candidates.  The $\sigma_{<V-I>}$ and $\sigma_{<B-V>}$ were about 0.02 mag in most cases, and 0.03 mag for $\sigma_{<V-I>}$ for all candidates --- about the same order of magnitude as the offsets themselves, suggesting that the color offsets are insignificant.  They are also in the opposite direction compared to the Milky Way cluster color offsets: SDSS is bluer while the Milky Way is redder.  If we look at only the 36 confirmed clusters, we find that they do have a $V-I$ color offset in the positive direction (where SDSS is redder than our colors), but smaller than we find in our comparison to the Milky Way --- only 0.02 mag, and about equal to $\sigma_{<V-I>}$.  We find a slightly larger color offset in $B-V$: 0.03 magnitudes bluer than our colors, with a $\sigma_{<B-V>}$ of about 0.02 mag.  A bluer $B-V$ affects the position of the clusters on the color-magnitude diagram in the same sense as a redder $V-I$ would, but the scatter in the SDSS colors is large, so it is difficult to confirm any significant systematic error in our colors on the basis of the SDSS comparison.

The ``fair'' candidates --- the objects having GC-like colors but large sizes --- are fainter and redder on average than the ``good'' and confirmed GC candidates, as one might expect for a sample heavily contaminated by galaxies.  The ``fair'' candidates have $<$($B-V$)$>$ = 1.07 mag and $\sigma_{(B-V)}$ = 0.10 mag, compared to $<$($B-V$)$>$ = 0.99 mag and $\sigma_{(B-V)}$ = 0.1 mag for the ``good'' and confirmed candidates.  The $V-I$ color is similarly offset: the fair candidates have $<$($V-I$)$>$ = 1.34 mag and $\sigma_{(V-I)}$ = 0.11 mag, while the ``good'' and confirmed candidates have $<$($V-I$)$>$ = 1.25 mag and $\sigma_{(V-I)}$ = 0.17 mag.  The mean $V$ of the ``fair'' candidates is 22.02 mag with a dispersion of 0.81 mag, while the ``good'' and confirmed candidates have a mean $V$ of 20.90 mag with a dispersion of 1.4 mag.

With colors in three different bands, we can construct the reddening-free parameter \citep{vdb67} $Q_{BVI}$ in the manner of \citet{bar00} in order to obtain a rough estimate of the overall reddening of the GCs.  The reddening-free parameter is defined as follows:
\begin{equation}
Q_{BVI} \equiv (B-V) -\frac{E_{B-V}}{E_{V-I}} (V-I) = (B-V)_0-\frac{E_{B-V}}{E_{V-I}} (V-I)_0.
\end{equation}
The ratio of $E_{B-V}$ to $E_{V-I}$ is determined using the \citet{ccm89} reddening relations with R = 3.1.  Performing a linear least-squares fit to $Q_{BVI}$ vs. ($B-V$)$_0$ relation for Milky Way GCs, we found 
\begin{equation}
Q_{BVI} = (0.882\pm0.079)(B-V)_0+ (0.573\pm0.014).
\end{equation}

The relation had a correlation coefficient of 0.57, indicating that the $Q_{BVI}$ provides a rather imprecise estimate of the true color of a GC.  One might expect such imprecision in dereddening for a population of clusters with a wide range of ages and metallicities.

For our 85 spectroscopically confirmed GCs, the mean $Q_{BVI}$ value was 0.18, which gives a mean ($B-V$)$_0$ of 0.73 mag and an mean total $E(B-V)$ of 0.25 mag.  The maximum $E(B-V)$ for confirmed GCs is 0.53 mag.  The mean $V$-band extinction $A_V$ for confirmed GCs is 0.73 mag.  This $A_V$ value includes the foreground absorption, which is estimated to be 0.27 mag in \citet{sch98}.  Removing the foreground absorption gives a typical internal extinction value of $A_V = 0.46$ for M81 GCs projected onto the disk.  The mean ($V-I$)$_0$ for spectroscopically confirmed GCs is 0.85 mag.

The reddening corrections based on the $Q$-parameters are not sensitive to whether color differences in GCs are caused by reddening or metallicity.  Therefore, all metallicity information and nearly all color difference is lost when these corrections are applied to individual GCs.  These reddening estimates are, however, useful for estimating a dereddened GCLF and for estimating the typical total reddening, foreground plus internal, in M81.

For all 221 spectroscopically confirmed and ``good'' GCs, the mean $Q_{BVI}$ and intrinsic colors ($B-V$)$_0$ and ($V-I$)$_0$ were identical to the results for the confirmed clusters alone.  The mean total $A_V$ was 0.77 mag, slightly higher than for the confirmed clusters alone.

\subsection{Luminosity Function and Specific Frequency}
The luminosity histograms for all 221 good and spectroscopically confirmed clusters with rough dereddening estimates applied are shown in Figure 11.  (The number of actual good and confirmed GC candidates, not corrected for completeness, is shown in each bin.)  The expected GCLF turnover magnitude in $V$, assuming an absolute magnitude of $-7.5$ and the \citet{fre01} distance modulus of 27.67, is 20.17 mag.  This value appears to be close to the turnover in the dereddened GCLF histogram.

To estimate the turnover magnitude and width of the GCLF, we use the MAXIMUM program written by J. Secker and described in \citet{sec93}.  We fit the Student's t5 distribution to our GCLF, which \citet{sec93} claim is superior to a Gaussian as a GCLF model.  Along with the list of magnitudes in $V$ of GCs, we also enter completeness information as a function of magnitude before fitting the GCLF.  The basic t5 fit is shown as the dotted line in Figure 10 with the $V$-band luminosity distribution.  The fit has been normalized by dividing each point in the function by the total probability distribution of the function sampled every 0.5 magnitudes (this total probability is 2), and by multiplying by the number of globular clusters (221).  The $V$ GCLF peak is 20.26 $\pm$ 0.13 mag, and the $\sigma$ of the GCLF is 1.49 $\pm$ 0.14.  The program failed to fit the peaks for the B and I luminosity distributions, but their turn-over magnitudes can be estimated from that of $V$ and the mean dereddened GC colors.  Adding the mean ($B-V$)$_0$ = 0.73 mag to the $V$ peak to estimate the $B$ peak gives $B_0$ = 20.99 mag, and subtracting the mean ($V-I$)$_0$ = 0.85 mag from the $V$ GCLF peak gives $I_0$ = 19.41 mag.  Both of these values are consistent with the apparent peaks of the $B$- and $I$-band histograms.  If we use the \citet{fer00} calibration for GCLF distance determination ($M_{V0}$ = --7.6 $\pm$ 0.25 mag), we find a distance modulus of 27.86 $\pm$ 0.28 mag for M81 and a linear distance of 3.73 Mpc.  This distance modulus is similar to the Cepheid-derived distance moduli of 27.8 and 27.67 determined by \citet{fre94} and \citet{fre01} respectively, as well as the distance modulus of 27.92 determined by \citet{mag01} using planetary nebulae.

If we assume the \citet{fre01} distance modulus of 27.67, the turnover magnitude of the M81 GCLF is M$_V$ = -7.41 $\pm$ 0.15 mag.  This is very similar to the MW turnover M$_V$ = -7.4 \citep{har01}, the M104 turnover M$_V$ = -7.60 $\pm$ 0.06 \citep{spi06}, and the M31 turnover M$_V$ = -7.65 $\pm$ 0.16 calculated using the \citet{bar01} turnover apparent magnitude and the \citet{jos03} distance to M31.  The dispersion in these four early-type spiral galaxies' GCLF turnover magnitudes is 0.13 mag.

Taking our 221 probable GCs as an estimate of the total number of GCs in M81, we find a GC specific frequency \citep{har81} $S_N$ = 1.13, similar to the typical $S_N$ $\sim$ 1 for early-type spiral galaxies (S0/a-Sbc).  If we calculate the specific frequency in a more traditional way, by doubling the bright half of the luminosity function ($V_{0}\leq$ 20.26 mag), we find 116 bright GC candidates and thus a total estimate of 232 GCs.  This gives an $S_N$ value of 1.18.  However, we are likely to be missing a substantial portion of the M81 GCS, particularly the halo clusters, due to the spatial limits of our survey.  Spatial coverage of our survey starts to be incomplete at slightly less than 7 kpc, and there is no coverage beyond 12 kpc.  Figure 12 shows the logarithm of the number density of GCs per square kpc as a function of projected galactocentric distance for our M81 sample (solid line), an M31 sample of 406 objects identified as spectroscopically confirmed GCs (tags ``1'' and ``9'') in the Revised Bologna Catalog of M31 globular clusters \citet{gal04}$^{3}$, and the MW \citep{har96}.  For the Milky Way, the projection used is $D_{proj} = (Y^2+Z^2)^{0.5}$, with the Y and Z coordinates taken from the \citep{har96} catalog and represent a coordinate along the direction of Milky Way rotation and above the Galactic plane, respectively.  Number densities were determined by counting the number of GCs in 1 kpc bins and dividing by the area of the 1 kpc annulus within which the GCs fall.  For M81, since the faint half of the GCLF is affected by potentially spatially dependent, luminosity-related incompleteness and contamination, the GCs brighter than the GCLF turnover were counted, and their numbers within each bin were doubled to estimate the M81 GC number density.  The last complete bin or annulus in the M81 GC sample is the 5-6 kpc bin; therefore the M81 GCS is expected to be complete within 6 kpc.  A simple way to estimate the number of missing halo GCs in M81 is to estimate the fraction of the GCS located within 6 kpc of the nucleus for the MW and M31, assume that the fraction of the M81 GCS within 6 kpc is the same as in these galaxies, and then divide by this fraction to get the total number of GCs.  There are an estimated 204 GCs (102 GCs in the bright half of the GCLF) within 6 kpc of the center of M81.  In the MW, GCs within a projected galactocentric distance of 6 kpc represent 62\% of the total sample (97 out of 157 GCs), so if the estimated 204 GCs within 6 kpc in M81 represent the same fraction as in the MW, M81 has an estimated 329 GCs.  This yields a GC specific frequency of 1.68.  The M31 sample has only 45\% of its GCs within 6 kpc; assuming this fraction for M81 yields a total sample of 453 GCs and a specific frequency of 2.31.

\footnotetext[3]{We use Version 4 of the catalog, revised December 2009, located at http://www.bo.astro.it/M31/.}

Further incompleteness in the GC sample would be expected to result from the rejection of too compact (rejected as possible stars) or too diffuse (``fair'') candidates.  There are only 6 ``fair'' candidates brighter than the GCLF turnover peak within 6 kpc, leading to an estimated 12 missing clusters within this radius (assuming all these bright inner ``fair'' candidates are GCs and not galaxies), or about 23-27 missing clusters at all projected radii.  And if 10\% of GCs were rejected as too compact, there would be 20 missing clusters within 6 kpc and $\sim$38-44 more missing clusters overall.  This would result in $\sim$61-71 more M81 GCs, or 446-524 total GCs.  A total GC population of 524 GCs would result in a specific frequency of 2.68.

Adding the ``fair'' candidates to the dereddened GCLF makes the $V$-band turnover magnitude 0.26 mag fainter than when they are excluded: $V_0$ = 20.52 $\pm$ 0.11.  The $\sigma$ of the $V$-band GCLF is slightly decreased, to 1.42 $\pm$ 0.10.  Assuming the \citet{fre01} distance to M81, the absolute turnover magnitude $M_V$ is reduced to $-7.15$ $\pm$ 0.13, considerably fainter the other massive spiral galaxies.  We would expect there to be at least a few genuine GCs among the ``fair'' candidates, since there are two confirmed GCs above the 3 $\sigma$ size limit, and at least a few diffuse GCs are commonly found in massive spiral galaxies.  However, there is a high probability that the ``fair'' candidate sample contains a substantial number of galaxies, and only spectroscopy could determine the true nature of these extended, red objects. 

\subsection{Sizes}
We used the ISHAPE software \citep{lar99} on individual 300$\times$300 pixel $V$-band GC images cut out from the V-band mosaic to estimate the core and half-light radii of our GC candidates.  We created an empirical PSF using 62 moderately bright stars ($V\sim$19-21) located in various parts of the mosaic image.  Although the {\it{HST}} ACS PSF differs according to location on the chip, we use a single composite PSF.  We justify this on the basis that (a) objects located where individual images in the mosaic overlap will have two different chip positions, and thus an exact chip position and corresponding PSF form may not exist for every object, and (b) we do not expect to obtain a high level of accuracy or precision in morphological parameters, given the crowded fields many GCs occupy.  We chose to fit fixed c = 30 elliptical King models to all candidates, to minimize any artificial variations in half-light (effective) radius that would result from an inaccurate choice of concentration parameter.  (The ``c'' used here is a non-logarithmic concentration parameter, equal to the tidal radius divided by the core radius.  A value of c=30 is typical for Milky Way globular clusters.)  Due to the crowding of many of the fields resulting from the clusters being mostly projected onto the galaxy disk, we do not have much confidence in our ability to estimate the structural parameters with great accuracy.  The FWHM, axis ratio, $\chi^{2}$ values, and upper and lower error bars in the FWHM were calculated by the ISHAPE program.  \citet{lar07} contains equations to convert FWHM into core radius and half-light radius assuming a circular profile; we estimate these parameters using the mean of the major and minor axis FWHM values.  We used the R$_{eff}$/FWHM conversion factor for King 30 from Table 3 of \citet{lar07}, based on Equations (9) and (10) of \citet{lar07}, to estimate the half-light radius.  We used the {\it{HST}} mosaic pixel scale of 0.05$\arcsec$ pixel$^{-1}$ and the \citet{fre01} distance to M81 to convert all sizes to parsecs.  The distance scale of our M81 {\it{HST}} images is 0.83 pc pixel$^{-1}$, or 1.21 pixel pc$^{-1}$.

Table 4 lists the IDs, candidate quality, $V$ magnitudes, projected galactocentric distances, and ISHAPE King model fit parameters of the objects, including FWHM, axis ratio, half-light radius, and uncertainty estimates in the half-light radius.  The uncertainty was estimated based on the average absolute values of the upper and lower error bars in the major and minor axis FWHMs, converted into parsecs, and added in quadrature to the standard deviation of half-light radius values determined using linear concentration parameters of 5, 15, 20, 30, 50, 75, and 100.  Uncertainties estimated in this way were typically around 40\%, rather high but probably reasonable given not knowing the concentration parameter for each cluster.  Figure 13 shows the half-light radius of ``good'' and spectroscopically confirmed GC candidates as a function of projected galactocentric distance and $V$ magnitude.  The typical sizes of M81 GCs appear to increase very slightly as a function of projected distance, although there are several relatively large GC candidates projected onto the inner regions of M81, and there is great scatter in sizes among fainter GCs.  If they are true GCs, many of these objects may be at large heights from the disk along the line-of-sight direction, and thus at small projected galactocentric distances but large absolute galactocentric distances.  Among some of the objects on the brightest end of the luminosity function, there is a hint of an upturn in half-light radius among brighter objects.  The median half-light radius of all confirmed + good GCs is 2.63 pc, smaller than the Milky Way median half-light radius of ~3 pc, but similar to the peak of the \citet{pea10} half-light radius distribution of mostly inner and intermediate-distance M31 GCs.

In Figure 13, it is apparent that there are two unusually large GCs among the spectroscopically confirmed objects, with half-light radii of about 12 pc and 19 pc.  They are too faint to be dwarf galaxies.  Fitting a linear concentration parameter c = 5 gives slightly smaller, but still large, sizes for them: about 10 pc and 17 pc.  The larger and brighter of the two has been noted by \citet{cft01} as being larger than their other candidates but clearly a cluster.  Their half-light radii may exceed those of other M81 GCs within their projected galactocentric distance, yet they are not as extremely extended as the 30 pc M31 GCs in \citet{hux05}.  There do exist 12 Milky Way GCs with half-light radii at or above 10 pc, most of them fainter than the GCLF turnover, suggesting that clusters this diffuse may be somewhat uncommon but not extraordinary.

A linear least-squares fit to all 85 spectroscopically confirmed clusters (including the one without a metallicity) gives only marginally significant evidence for a size gradient: a slope of 0.17 $\pm$ 0.13 pc per kpc of projected distance.  A fit to all 221 ``good'' and confirmed candidates gives stronger evidence for any GC size gradient: a slope of 0.36 $\pm$ 0.06 pc per kpc.  It is possible, however, that the faint end of the good candidate GCLF has significant contamination from background galaxies.  To minimize the background galaxy contamination problem, we perform a least-squares linear fit to the size versus projected distance relation for all confirmed candidates plus 40 ``good'' candidates brighter than $V$ = 21.07 (the faintest of the confirmed candidates).  GC candidates this bright should have a very low probability of being background galaxies.  Fitting only bright GC candidates, we find a size gradient of 0.26 $\pm$ 0.09 pc per kpc of projected galactocentric distance.  As we saw in Section 5.2, only 8\% of spectroscopically confirmed objects in our catalog of 419 GC candidates are non-GCs.  Assuming an 8\% contamination rate in the bright GC candidate sample, only about 3 of the 40 ``good'' candidates (and thus 3 out of 125 bright objects) are expected to be non-GCs.  Therefore, due to the combination of good sample size and low contamination, the bright sample of 125 objects would be expected to provide the most accurate size gradient as a function of galactocentric distance.

In our M81 spectroscopically confirmed sample, the MP clusters have a median half-light radius of 2.30 pc, and the MR clusters have a median half-light radius of 2.01 pc.  (The median half-light radius of all 85 confirmed clusters is 2.13 pc.)  The ratio of the median MR half-light radius to the median MP half-light radius is thus 0.87 --- the MR clusters are nearly as large as the MP ones.  Within 2 kpc of the nucleus, the median size of MR clusters is 70\% of the size of the MP clusters.  Since we do not have GC candidates at large projected radii, we cannot determine for certain whether MP and MR clusters would have the same size at large galactocentric distances.  Our most complete sample of GCs is within 7 kpc of the center of M81, representing 2.1 effective radii according to \citet{rc3}, while our total sample extends out to 3.6 effective radii.

If we compare the sizes of ``blue'' clusters ($V-I$ $\leq$ 1.18 mag, the median $V-I$ color of confirmed GCs) and ``red'' clusters ($V-I$ $>$ 1.18), which makes our results more directly comparable to those of more distant galaxies where spectroscopic metallicities are not a concern, we find that they are very similar in median half-light radius.  In the spectroscopically confirmed subsample, the median size of red clusters is slightly larger than that of blue clusters, by a factor of 1.06; in the ``good'' plus confirmed sample, red and blue clusters are identical in median half-light radius.  These results may be complicated by the fact that we do not correct for reddening because, as mentioned in Section 5.2, the Q-parameter-based reddening estimates do not preserve intrinsic color differences due to metallicity (and thus are only useful for correcting the average magnitude of the GC population).

Figure 14 shows the half-light radii of red and blue (as defined in the paragraph above) spectroscopically confirmed M81 GCs (left) and Milky Way GCs with reddening less than 0.5 to match the color and reddening distribution of our M81 sample (right) as a function of projected galactocentric distance.  (The projection for the Milky Way GCs is the same as used for Figure 12.)  Also shown are least-squares linear fits to the half-light radii as a function of projected distance.  In the Milky Way, GC radii increase as a function of absolute galactocentric distance  \citep{vdb91}.  With the projection we have chosen, this trend cannot be seen: a linear fit to the Milky Way objects shown gives a slope where size actually decreases very faintly with projected distance, but with no statistical significance: $-0.002$ $\pm$ $0.012$ pc per kpc .  The M81 spectroscopically confirmed GCs, however, do show somewhat of a trend of increasing size with projected distance.   As mentioned above, a linear least-squares fit to all 85 spectroscopically confirmed clusters (including the one without a metallicity estimate) gives only marginally significant evidence for a size gradient, 0.17 $\pm$ 0.13 pc per kpc.

The size distributions of MR and MP GCs in M81 appear for the most part similar, and a two-sample K-S test of the size distributions gives P = 0.698, lending statistical support to the similarity of the MR and MP size distributions.  In other galaxies, both early type \citep{kun98,kun99,lar01a,gom07} and spiral \citep{bar02,lar01b}, there is evidence of size differences between MR and MP GCs.  This size difference could be attributed to a projection effect \citep{lar03}: MP clusters at a given projected distance are more likely to be at large absolute distances from the nucleus than the MR clusters at the same projected distance, and the MP/MR size difference is simply a result of the correlations between distance and metallicity and between distance and half-light radius.  However, \citet{jor04} notes the possibility of a genuine, direct correlation between metallicity and half-light radius, which may be attributable to the effects of mass segregation and the different luminosities of MR and MP stars of the same mass.  More recent studies such as \citet{mad09}, \citet{har09}, and \citet{har10} and references therein find that red GCs still tend to be smaller than blue GCs in massive galaxies even at a given galactocentric distance, suggesting that an intrinsic correlation between metallicity and half-light radius is at least partially necessary to explain the size differences found in red and blue clusters.  \citet{har09} suggests that, along with mass segregation effects, such a correlation could be caused by the clouds from which red GCs form contracting to a smaller size than those from which blue GCs form, due to the greater cooling effect from the heavy elements.  The MR/MP size differences in M81 could be attributed to any of these effects.  Also interesting is that the size difference in M81 is small (about 13\%) between MR and MP, compared to the 20-25\% often found in elliptical galaxies, and cannot be seen in a red-blue separation (although this can be blamed on the reddening, which is hard to correct for indivudually in the clusters, obscuring the color-metallicity relation).  \citet{har10}, \citet{mas10} and references therein suggest that the size difference between MP and MR or red and blue GCs is smaller in disk galaxies of type S0/a and later than in elliptical galaxies, and the size increase as a function of projected galactocentric distance appears to be larger in said disk galaxies than for elliptical galaxies, both for reasons not yet understood.  In \citet{mas10}, the environments in which such galaxies are found have been suggested to play a possible role: elliptical galaxies are more likely to be found in the richest parts of dense clusters than spiral galaxies, and these environments may reduce the effects an individual galaxy's potential well have on the relative sizes of globular clusters, thus reducing the size-galactocentric distance relation.  Our evidence, pointing to only small size differences between MR and MP or red and blue clusters and, except in the pure spectroscopically confirmed sample, a substantial size-galactocentric distance trend, appears consistent with these recent findings for spiral galaxies.

\section{Summary}

We have performed {\it{HST}} ACS $BVI$ photometry and ISHAPE King model fitting of 419 GC candidates with $V \lesssim$ 23, and classified them according to color and size limits defined by confirmed GCs.  We found 136 non-spectroscopically-confirmed objects that fit the typical color and size ranges of confirmed GCs.  When the ``good'' unconfirmed GC candidates were combined with the 85 confirmed GCs in our sample, we had a full sample 221 highly probable GCs.  ``Good'' GC candidates had a centrally concentrated spatial distribution similar to that of the confirmed GCs.  The dereddened $V$-band GCLF peak was $V_0$ = 20.26 $\pm$ 0.13, yielding an estimated distance modulus of 27.86 $\pm$ 0.28 assuming $M_{V0} = -7.6$ \citep{fer00}.  The half-light radii of GC candidates appeared to increase slightly with increasing galactocentric distance, and confirmed MR GCs were only about 10\% smaller than confirmed MP GCs.  The luminosity distribution was similar to other spiral galaxies such as the Milky Way and M31, and the sizes were similar to those of M31 GCs with small to intermediate projected galactocentric distances.

\section{Acknowledgments}

This research was supported by the NASA grant GO-10250 and GO-10584 from the Space Telescope Science Institute, the National Science Foundation grant AST 05077229, and the NASA grant NNG06GD33G from the GALEX Guest Observers Program.  We would also like to thank the staff of the Harvard College Observatory for their support.  Astrophysics research in Crete is supported by grant ASTROSPACE 206464.  A.Z. acknowledges support from the Marie Curie Reintegration Grant 224878.  J.N. gratefully acknowledges support from the Chilean
Centro de Excelencia en Astrof\'i sica y Tecnolog\'ias Afines (CATA).

\begin{figure}
\plotone{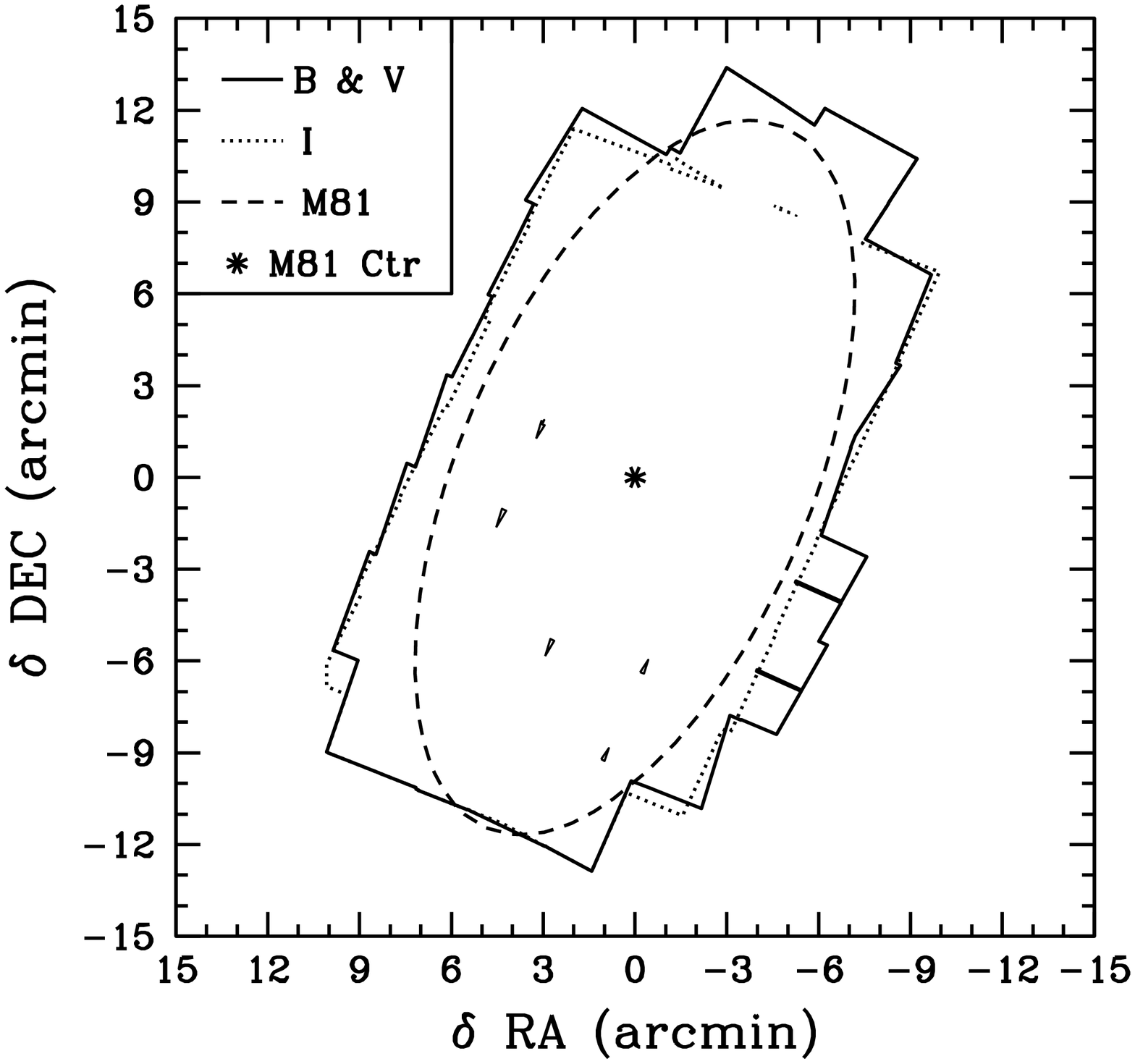}
\caption{Area covered by the {\it{HST}} mosaics.  The center of M81 is marked with a star, and the disk (25 mag arcsec$^{-2}$) is denoted with a dashed line.}
\end{figure}
\clearpage


\begin{figure}
\plotone{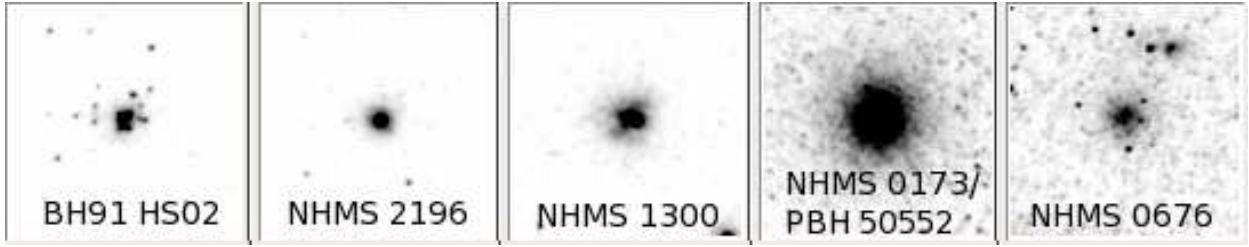}
\caption{5$\arcsec \times$5$\arcsec$ (86.4 pc $\times$ 86.4 pc) $V$-band image of BH91 HS-02 (left), a spectroscopic ``GC'' rejected from our catalog due to very blue Source Extractor $B-V$ and $V-I$ colors, compared to 4 other spectroscopic GCs which fit our GC color criteria.}
\end{figure}
\clearpage


\begin{figure}
\plotone{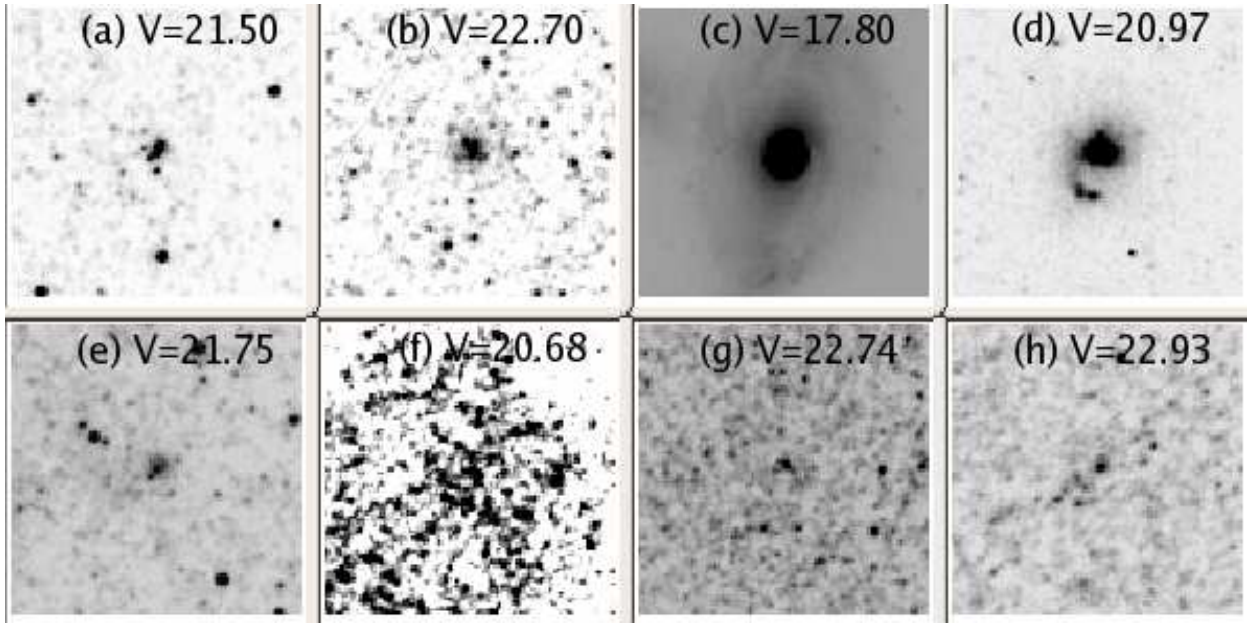}
\caption{5$\arcsec \times$5$\arcsec$ (86.4 pc $\times$ 86.4 pc) $V$-band images of objects visually rejected as GC candidates after passing the Source Extractor color, magnitude, and CLASS\_STAR cuts.}
\end{figure}
\clearpage


\begin{figure}
\plotone{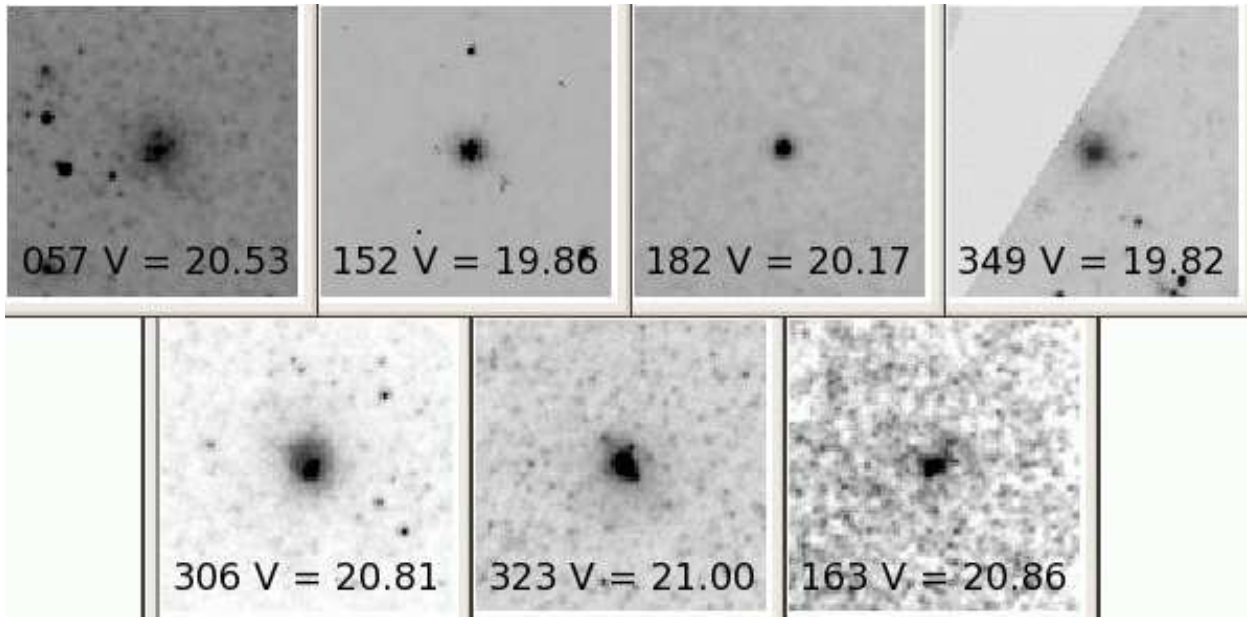}
\caption{5$\arcsec \times$5$\arcsec$ (86.4 pc $\times$ 86.4 pc) $V$-band images of the seven objects spectroscopically determined to not be GCs in \citet{nan10b}.  Top: four H {\small{II}} regions.  Bottom: three background galaxies.}
\end{figure}
\clearpage


\begin{figure}
\plotone{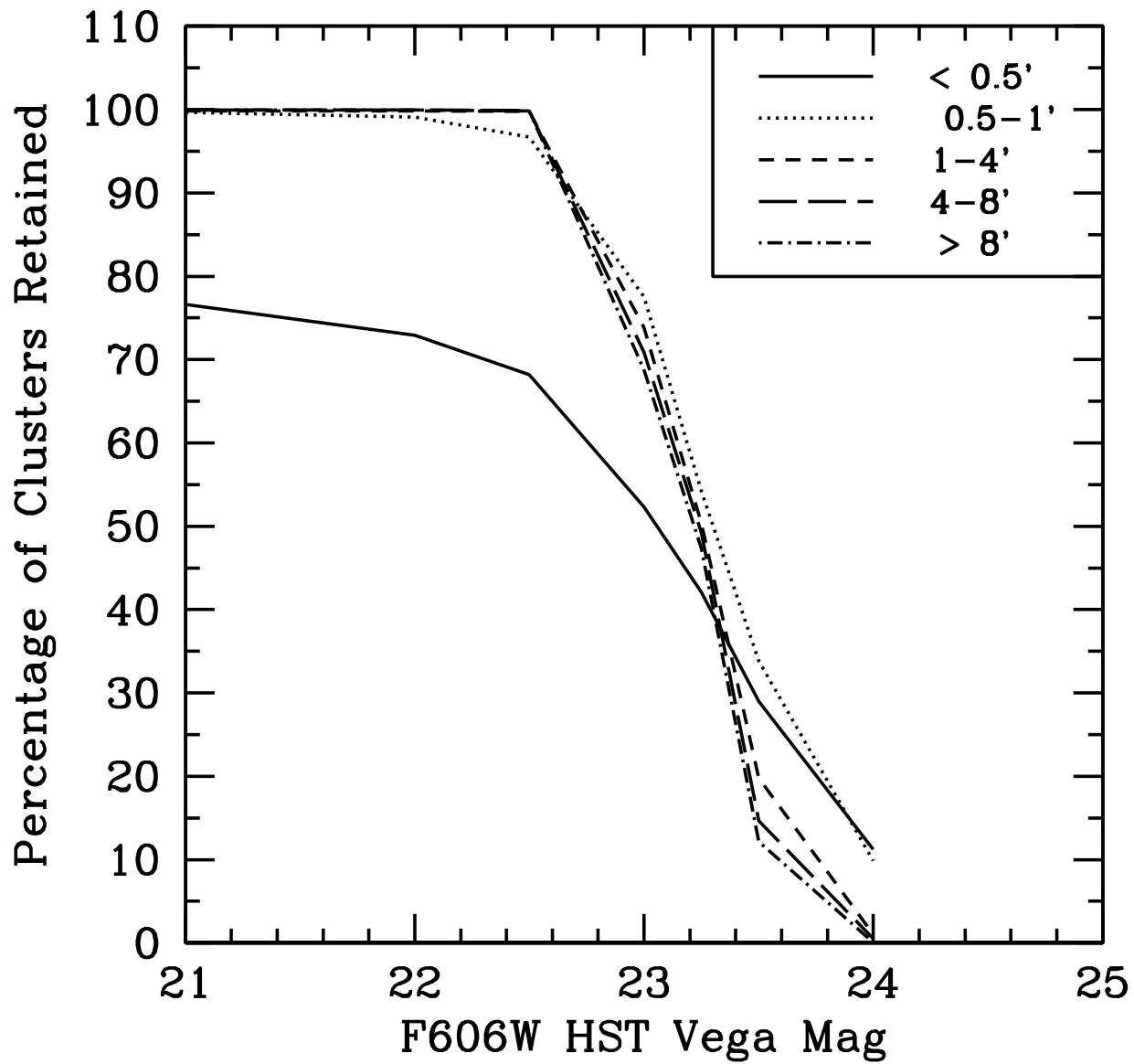}
\caption{Completeness as a function of magnitude for F606W VEGA magnitude (untransformed V), not accounting for potential losses due to surface brightness or compactness.  Different lines represent different projected galactocentric radius bins as specified in the plot legend.}
\end{figure}
\clearpage

\begin{figure}
\plottwo{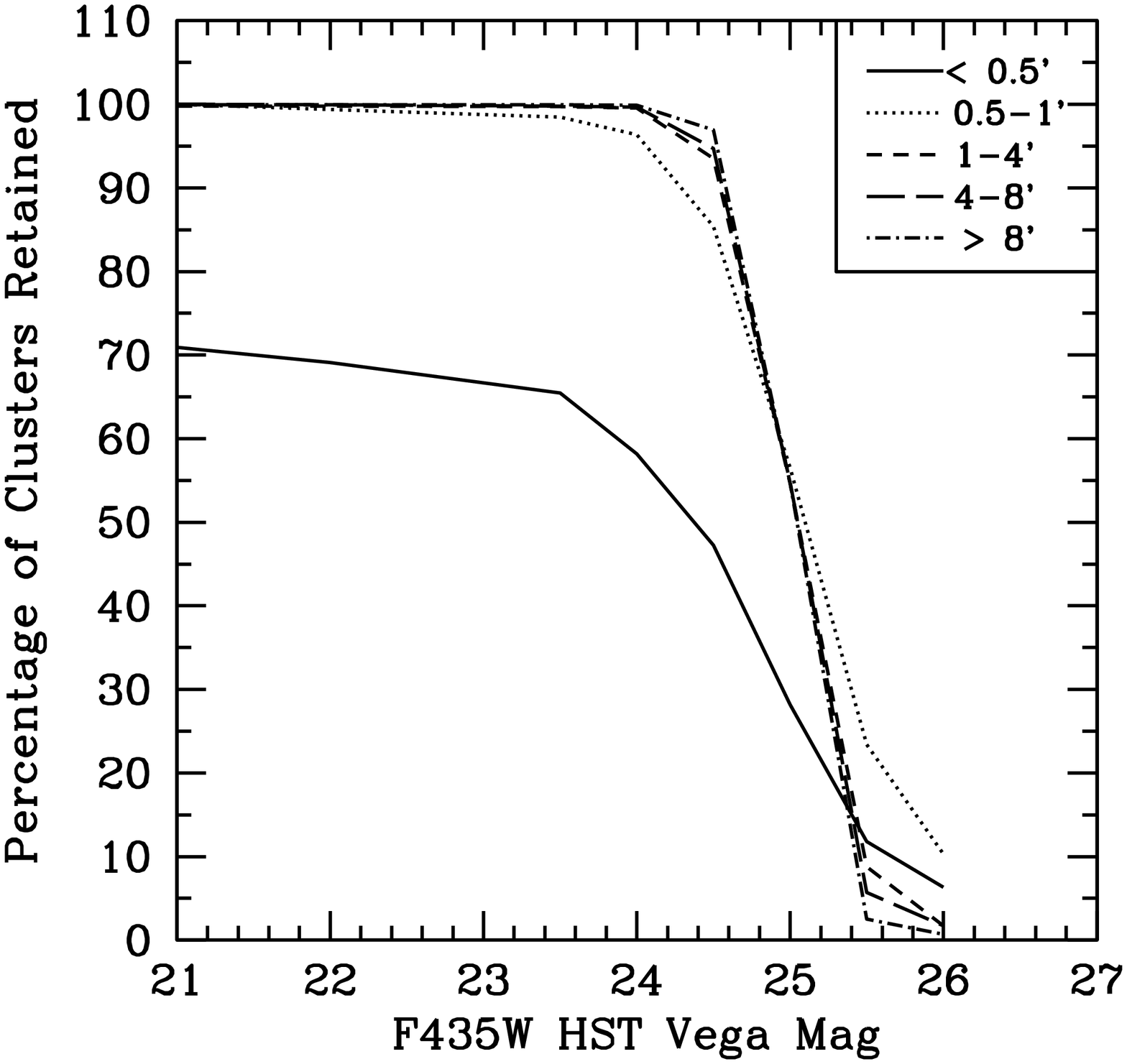}{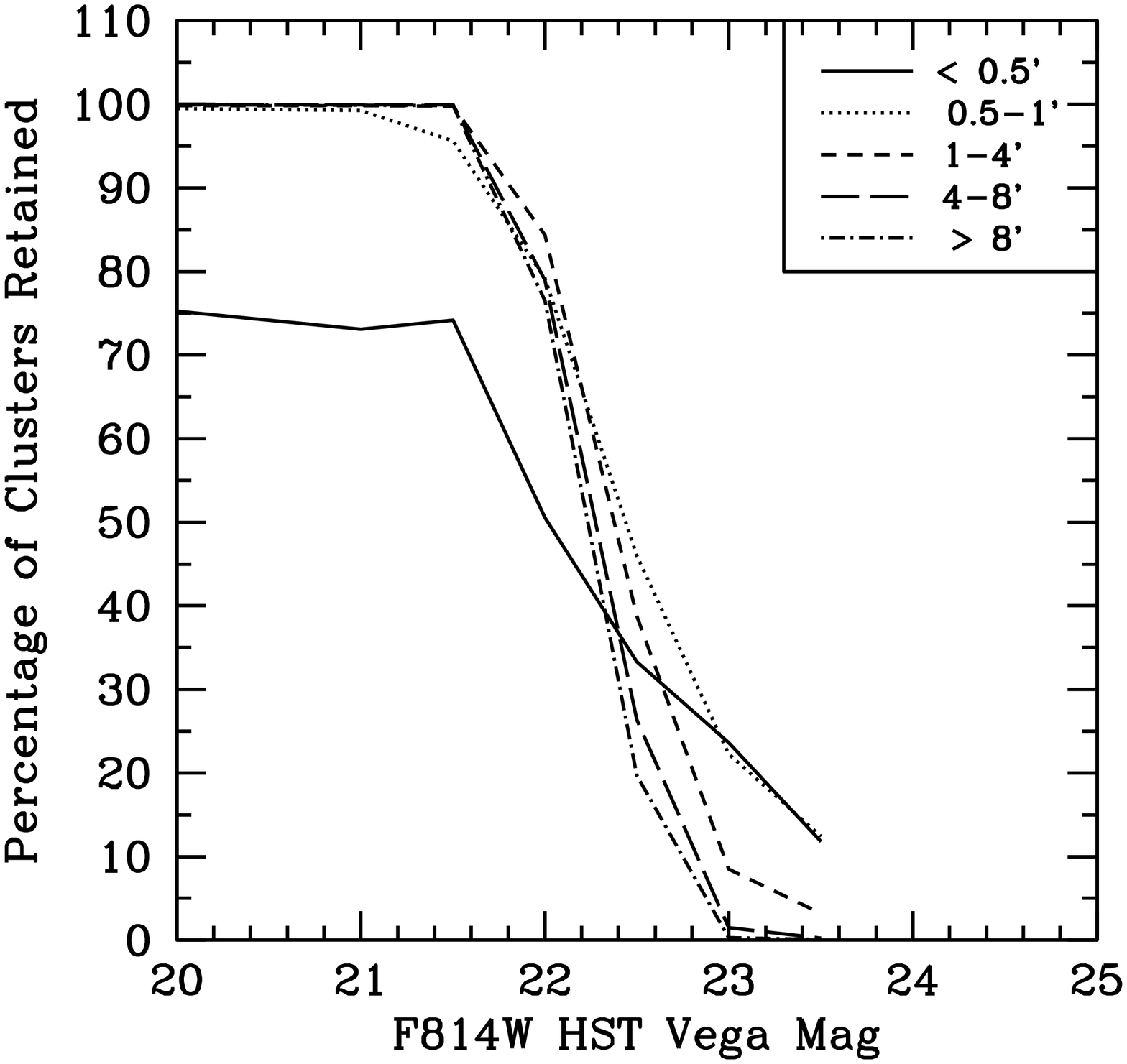}
\caption{Completeness as a function of magnitude for F435W (untransformed $B$) and F814W (untransformed $I$),  not accounting for potential losses due to surface brightness or compactness.  Different lines represent different projected galactocentric radius bins as specified in the plot legend.}
\end{figure}
\clearpage

\begin{figure}
\plotone{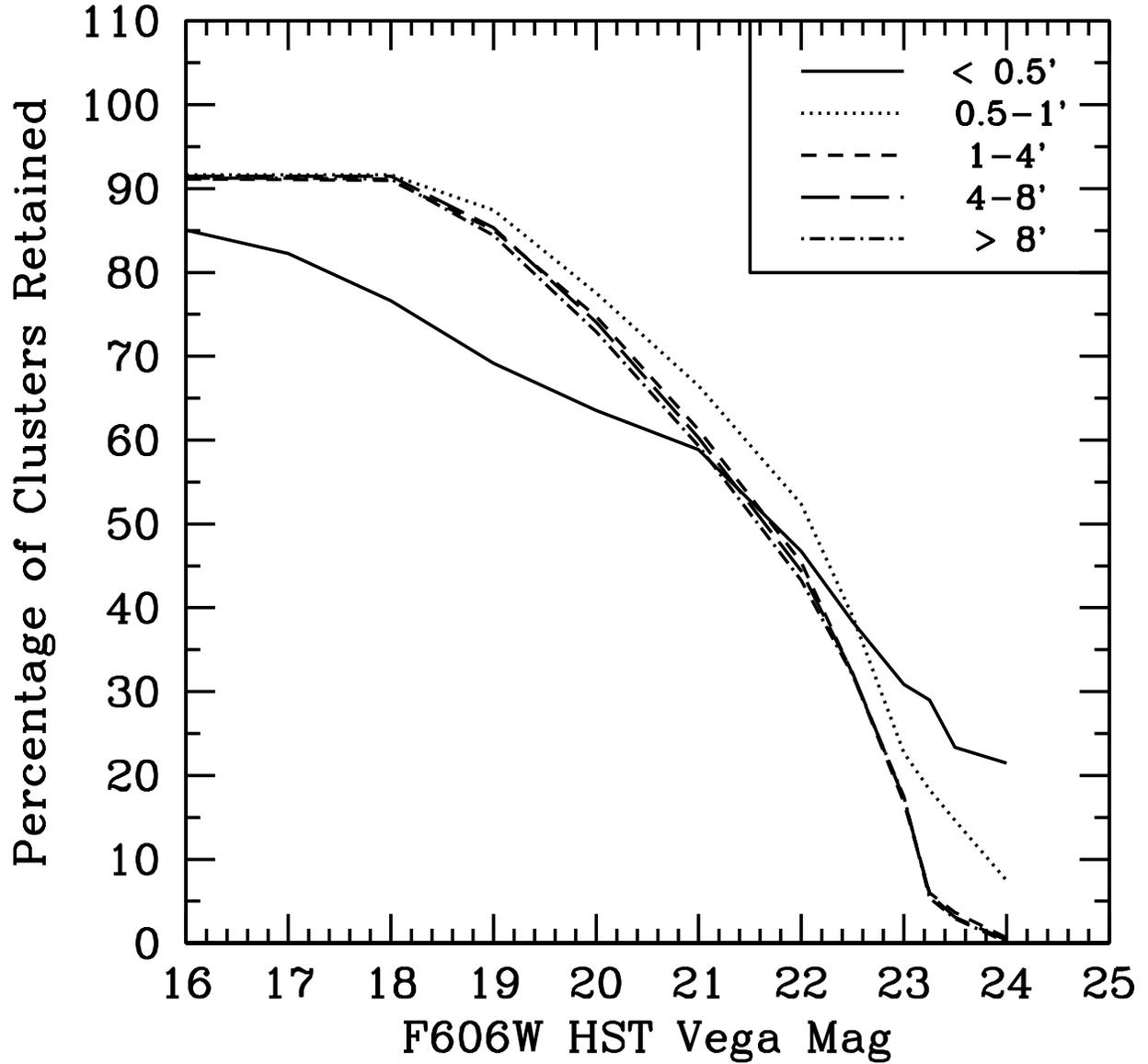}
\caption{Completeness vs. magnitude for King models of high-latitude Milky Way GCs, showing the initial loss due to compactness as well as the gradual losses due to low surface brightness.  Different lines represent different projected galactocentric radius bins as specified in the plot legend.}
\end{figure}

\clearpage

\begin{figure}
\plotone{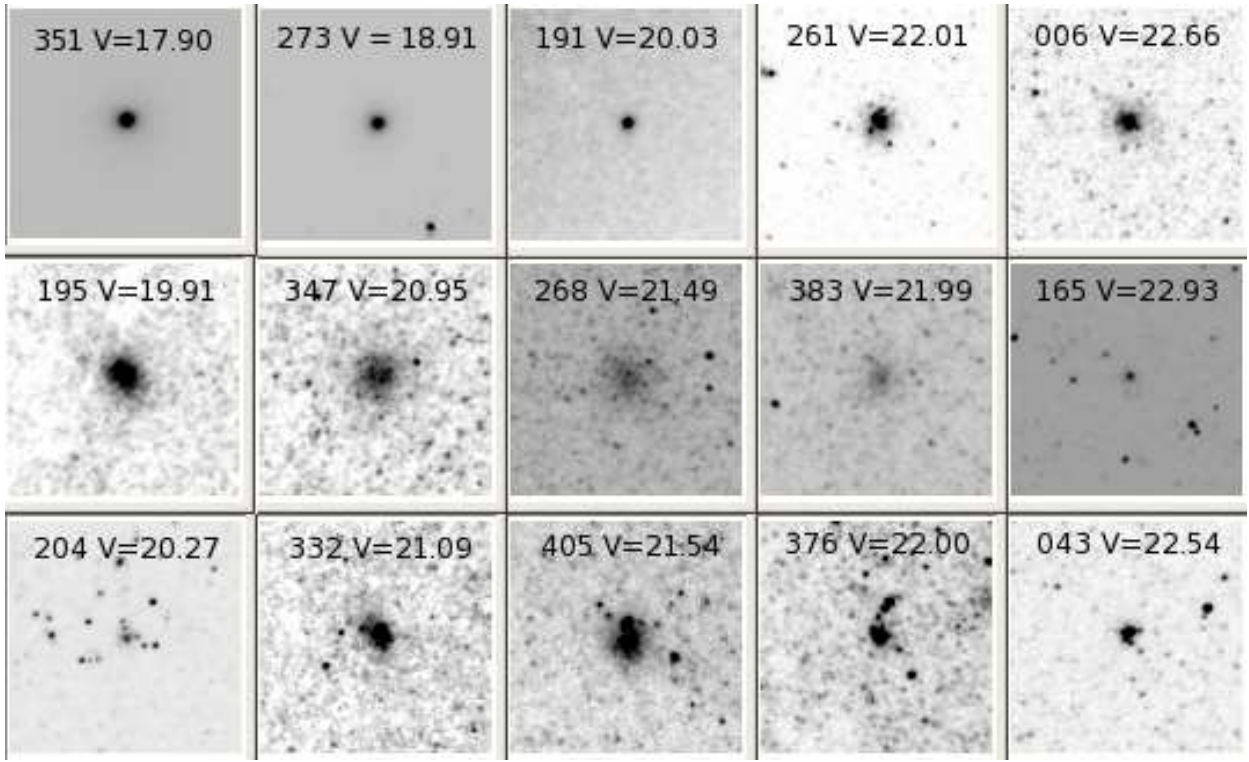}
\caption{5$\arcsec \times$5$\arcsec$ $V$-band images of''good'' (top row), ``fair'' (middle row), and ``poor'' (bottom row)  GC candidates.}
\end{figure}
\clearpage


\begin{figure}
\plotone{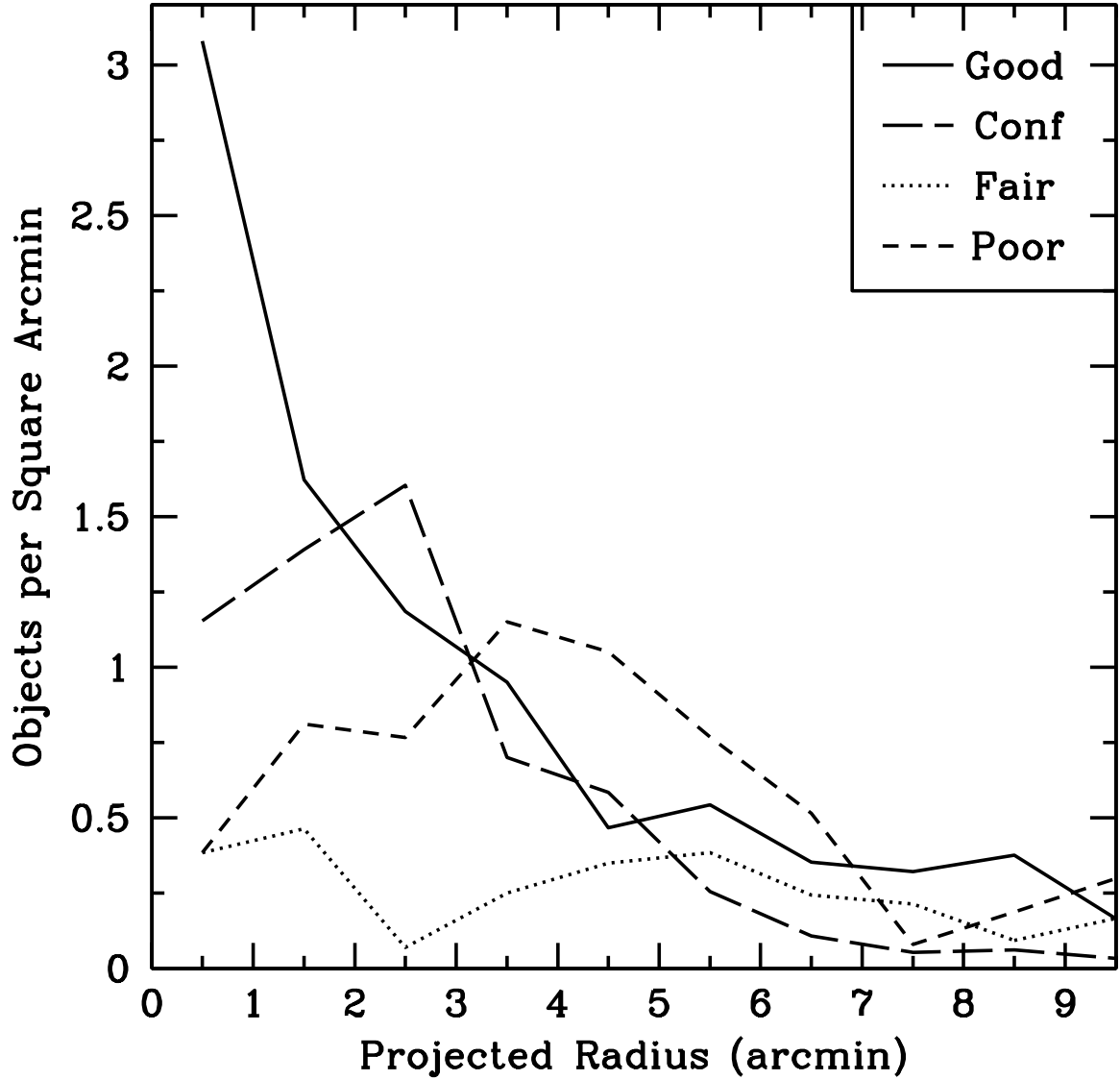}
\caption{Density of objects per square arcminute as a function of projected radius for ``good'' (solid line), spectroscopically confirmed (long-dashed line) ``fair'' (dotted line), and ``poor'' (short-dashed line) GC candidates.}
\end{figure}
\clearpage


\begin{figure}
\plotone{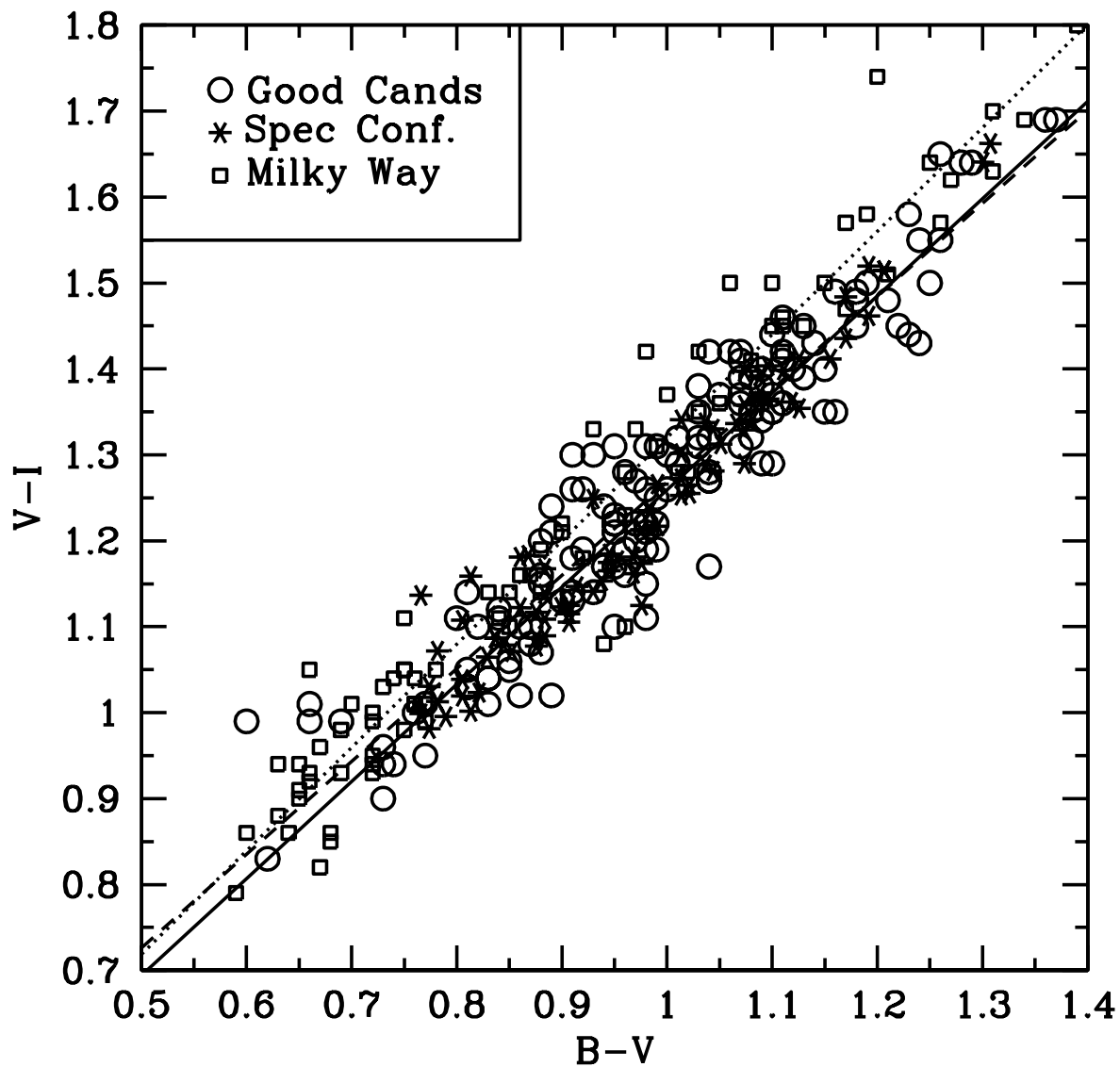}
\caption{Color-color plot for all good and spectroscopically confirmed GC candidates.  The solid line represents the color-color relationship for the spectroscopically confirmed M81 GCs, the dashed line is the color-color relationship for ``good'' but unconfirmed M81 GCs, and the dotted line is the color-color relationship for Milky Way GCs.  All M81 GC and GC candidate colors are derived from IRAF aperture photometry.}
\end{figure}
\clearpage

\begin{figure}
\plotone{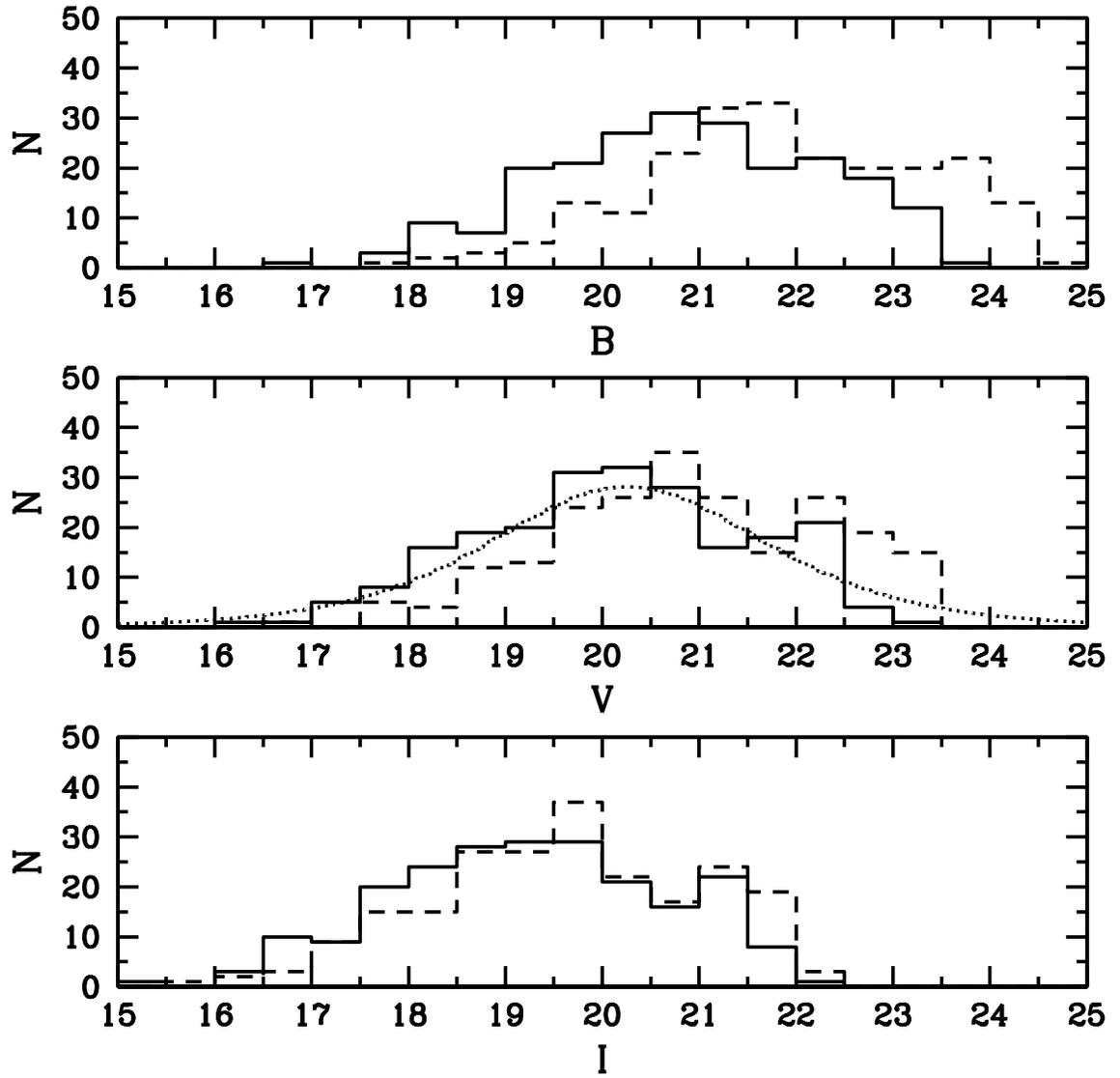}
\caption{GC luminosity histograms in $B$ (top), $V$ (middle), and $I$ (bottom) for confirmed and ``good'' candidates, uncorrected for completeness.  Dashed lines show luminosity distributions uncorrected for reddening.  The dotted line for $V$ shows the t5 fit to the $V$-band luminosity function.}
\end{figure}
\clearpage


\begin{figure}
\plotone{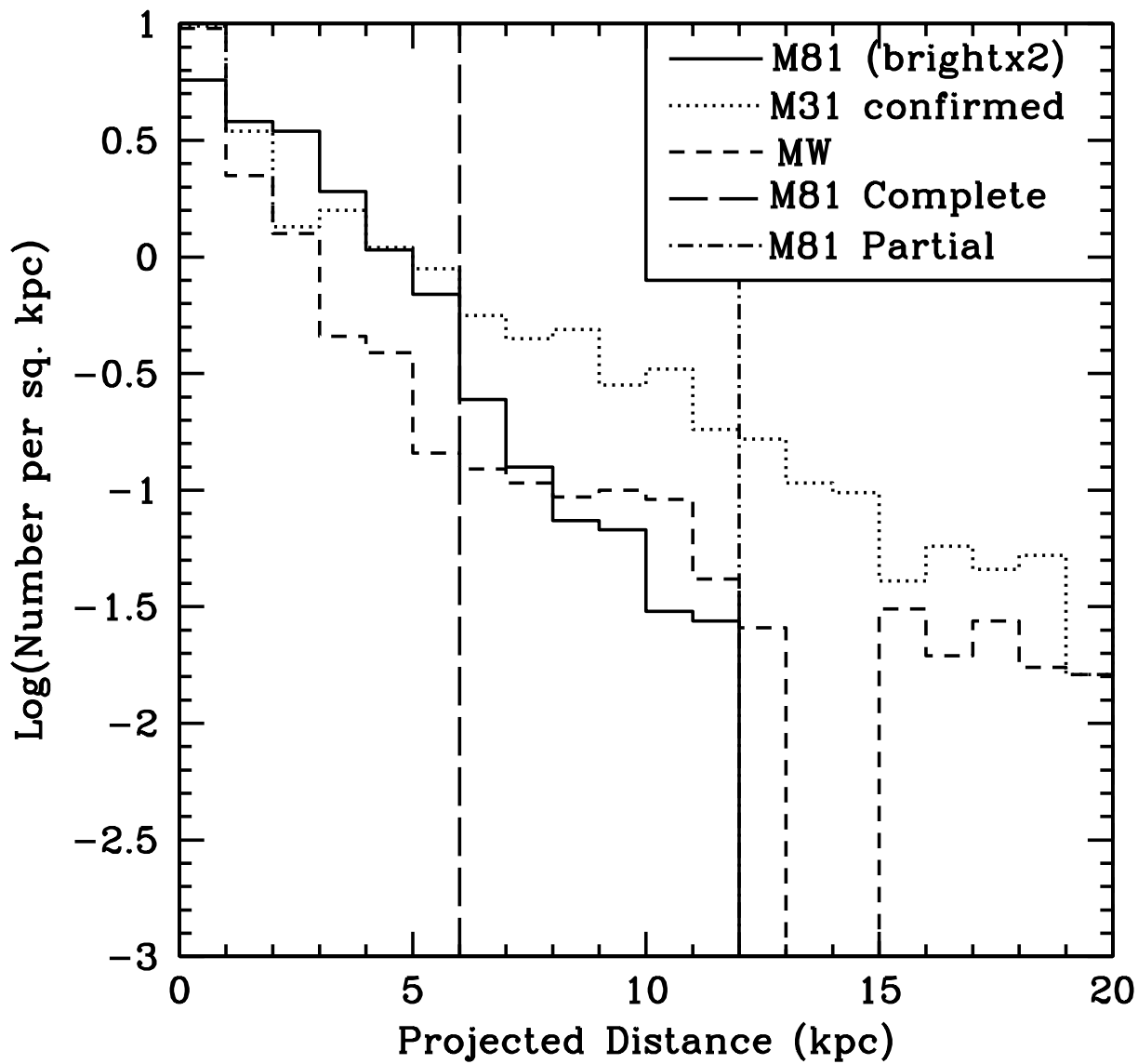}
\caption{Logarithm of the number of GCs per square kpc as a function of projected galactocentric distance for M81 (solid line), M31 (dotted line), and the MW (short-dashed line).  Numbers of GCs were determined for 1 kpc bins.  The long-dashed line at 6 kpc represents the limit within which the M81 GCS is expected to be complete, and the dash-dot line at 12 kpc represents the limit within which partial sampling of the M81 GCS is available.}
\end{figure}
\clearpage

\begin{figure}
\plottwo{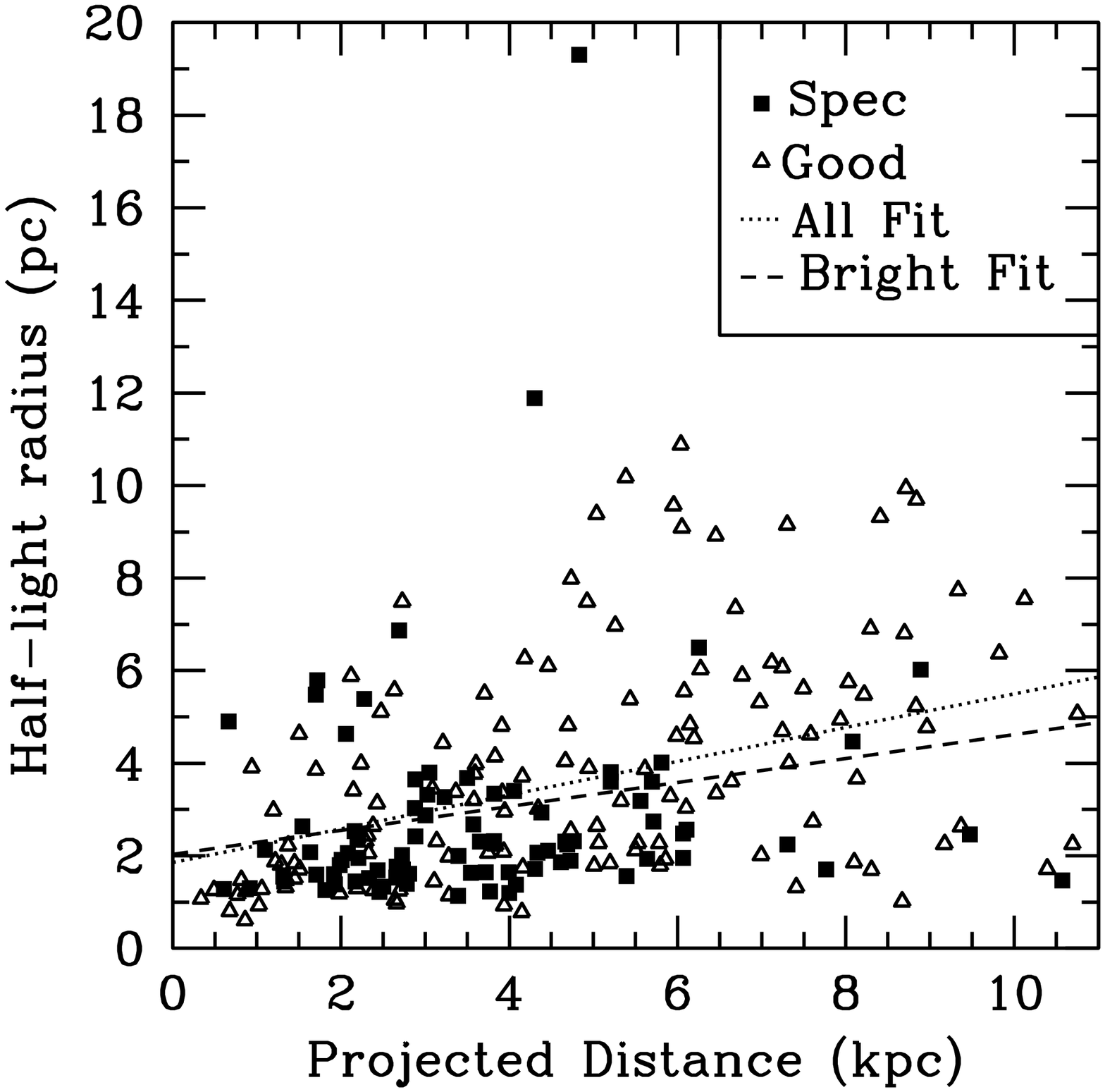}{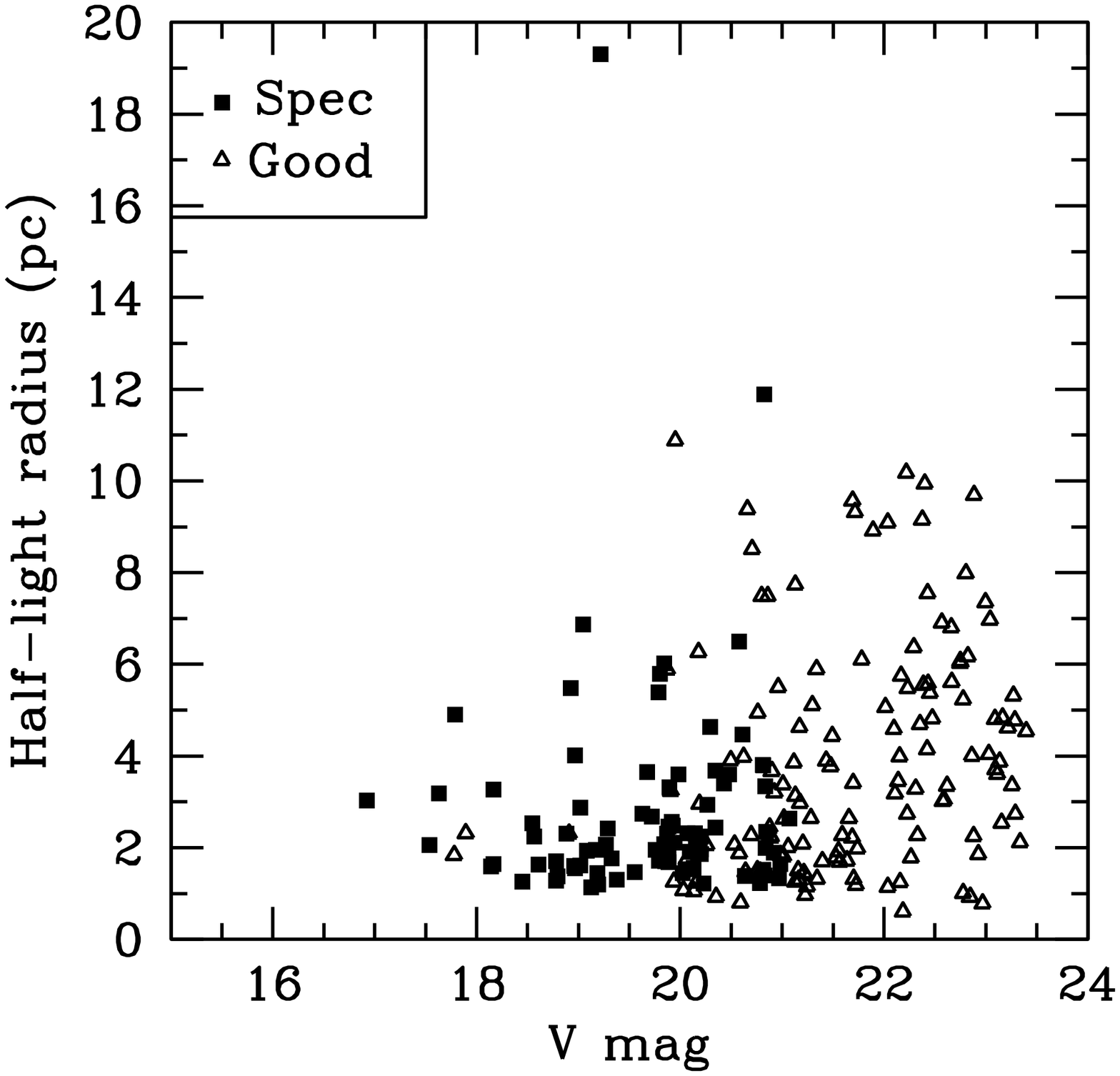}
\caption{Half-light radius as a function of projected distance (left) and $V$ magnitude (right) for spectroscopically confirmed and ``good'' GC candidates.  The dotted line is the half-light radius vs. projected distance relation for all ``good'' and spectroscopically confirmed GCs, and the dashed line is for only the spectroscopically confirmed clusters and those non-confirmed objects as bright as or brighter than the faintest confirmed cluster.}
\end{figure}
\clearpage


\begin{figure}
\plottwo{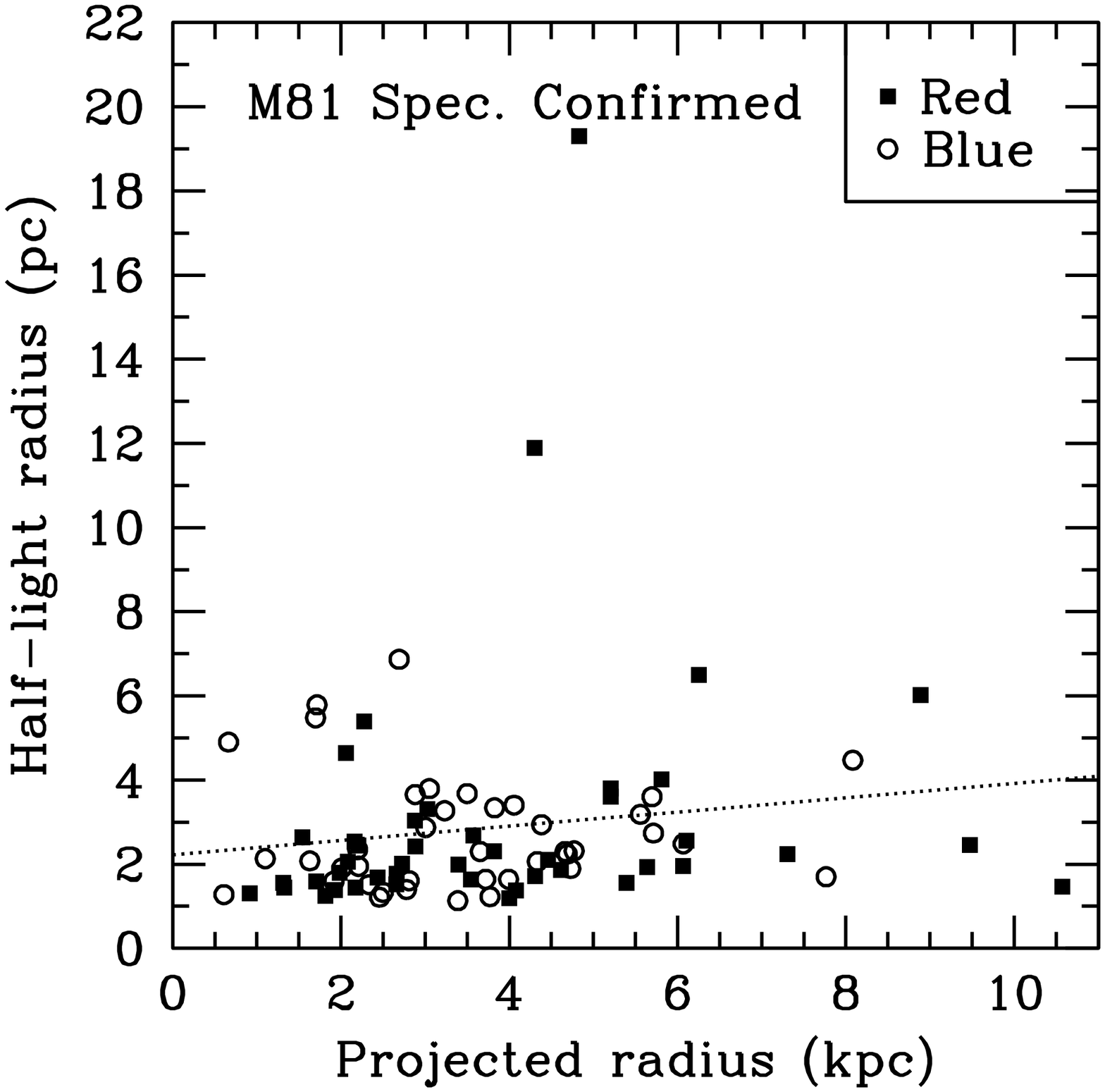}{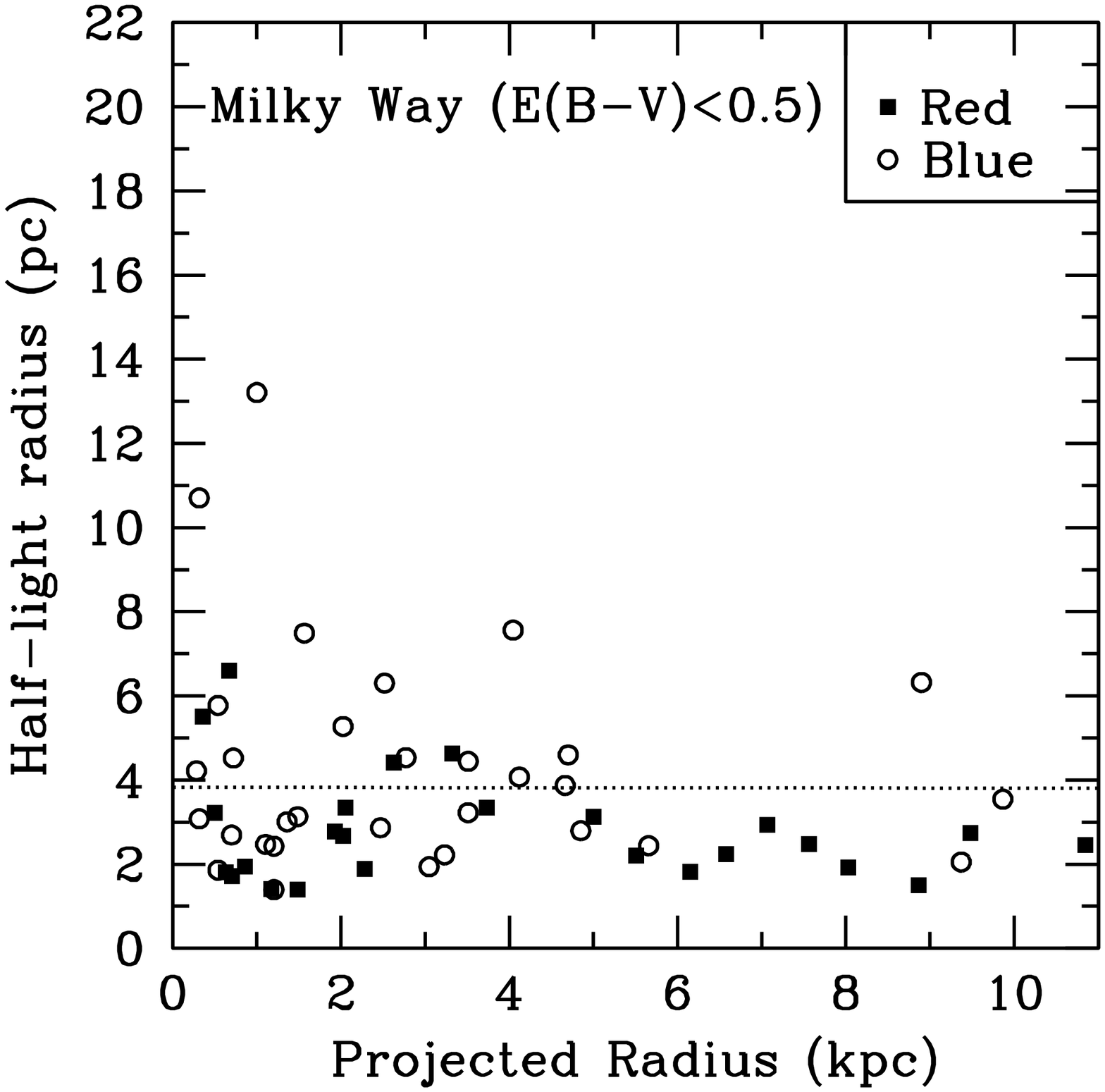}
\caption{Half-light radius as a function of projected distance for MR and MP GCs in M81 and the Milky Way.  ``Blue'' clusters are those with $V-I$ $\leq$ 1.18, the median color of M81 clusters, and red clusters are those with $V-I$ $>$ 1.18.  The colors of Milky Way clusters are not corrected for reddening, and only those with $E(B-V)$ $<$ 0.5 are shown, to make the color samples more comparable to those of M81.  The dotted lines are the half-light radius vs. projected distance relations.}
\end{figure}
\clearpage

\begin{deluxetable}{ccccccc}



\tablenum{1}
\tablewidth{425pt}
\tablecaption{Exposure Information}


\tablehead{\colhead{Proposal} & \colhead{ID} & \colhead{RA Ctr.} & \colhead{Dec Ctr.} & \colhead{Filter} & \colhead{Date} & \colhead{T$_{exp}$} \\ 
\colhead{} & \colhead{} & \colhead{(hours)} & \colhead{(degrees)} & \colhead{} & \colhead{} & \colhead{} } 

\startdata
10584 & Field 1   & 09 54 16.49   & 69 13 42.3   & F435W & 2006/03/23  & 1565 \\
10584 & Field 2  & 09 54 52.27  & 69 14 54.3  & F435W & 2006/12/31  & 1565 \\
10584 & Field 3   & 09 54 09.18   & 69 09 49.5   & F435W & 2005/12/10  & 1200 \\
10584 & Field 4   & 09 54 41.78   & 69 11 06.7   & F435W & 2005/12/06  & 1200 \\
10584 & Field 5  & 09 55 13.52  & 69 12 25.1  & F435W & 2006/03/22  & 1200 \\
10584 & Field 6   & 09 55 46.09   & 69 13 42.4   & F435W & 2006/03/23  & 1200 \\
10584 & Field 7   & 09 54 17.45   & 69 08 27.4   & F435W & 2005/12/15  & 465 \\
10584 & Field 7   & 09 54 22.89   & 69 06 56.8   & F435W & 2006/03/24  & 1200 \\
10584 & Field 8   & 09 54 55.47   & 69 08 14.2   & F435W & 2006/03/22  & 1200 \\
10584 & Field 9   & 09 55 11.57   & 69 08 49.8   & F435W & 2006/03/22  & 465 \\
10584 & Field 9  & 09 55 28.04  & 69 09 31.5  & F435W & 2006/03/20  & 1200 \\
10584 & Field 10   & 09 56 00.66   & 69 10 49.0   & F435W & 2006/03/25  & 1200 \\
...\\
\enddata

\end{deluxetable}
\clearpage


\begin{deluxetable}{ccccccc}



\tablenum{2}
\tablecaption{Basic Object Information}
\setlength{\tabcolsep}{0.065in}


\tablehead{\colhead{ID} & \colhead{RA} & \colhead{Dec} & \colhead{\citet{nan10a} ID} & \colhead{PR95 ID} & \colhead{CFT01 ID} & \colhead{Spec conf.} \\ 
\colhead{} & \colhead{(hours)} & \colhead{(deg)} & \colhead{} & \colhead{} & \colhead{} & \colhead{} } 

\startdata
1 & 09:53:52.45 & 69:08:56.96 & -- & -- & -- & -- \\
2 & 09:53:59.68 & 69:08:35.95 & 20 & -- & -- & -- \\
3 & 09:54:01.52 & 69:10:56.14 & 23 & -- & -- & -- \\
4 & 09:54:04.97 & 69:09:18.80 & 34 & 51040 & -- & (c) \\
5 & 09:54:11.83 & 69:08:46.51 & -- & -- & -- & -- \\
6 & 09:54:15.11 & 69:06:48.76 & 56 & -- & -- & -- \\
7 & 09:54:15.34 & 69:08:03.73 & 59 & -- & -- & -- \\
8 & 09:54:15.54 & 69:11:30.60 & 60 & -- & -- & -- \\
9 & 09:54:19.17 & 69:11:41.48 & 82 & -- & -- & -- \\
10 & 09:54:20.44 & 69:07:14.48 & 86 & -- & -- & -- \\
...\\
\enddata


\tablecomments{Objects from \citet{sch02} (b) and \citet{bro91} (d) have their object numbers in those catalogs specified in Column 7, and objects from \citet{nan10b} (c) that are not GCs have their object type specified in Column 7.}

\tablerefs{(a) = \citet{pbh95}; (b) = \citet{sch02}; (c) = \citet{nan10b}; (d) = \citet{bro91}.}

\end{deluxetable}
\clearpage


\begin{deluxetable}{ccccccccccc}



\tablenum{3}
\setlength{\tabcolsep}{0.05in}
\tablecaption{Photometry of GC Candidates}


\tablehead{\colhead{ID} & \colhead{$V$} & \colhead{$\sigma_V$} & \colhead{$B-V$} & \colhead{$\sigma_{B-V}$} & \colhead{$V-I$} & \colhead{$\sigma_{V-I}$} & \colhead{Apert.} & \colhead{FWHM} & \colhead{Flag} & \colhead{Qual.} \\ 
\colhead{} & \colhead{mag} & \colhead{mag} & \colhead{mag} & \colhead{mag} & \colhead{mag} & \colhead{mag} & \colhead{$\arcsec$} & \colhead{pix} & \colhead{} & \colhead{} } 

\startdata
1 & 22.33 & 0.02 & 1 & 0.03 & 1.57 & 0.03 & 1 & 2.53 & -- & poor \\
2 & 22.71 & 0.02 & 0.81 & 0.03 & 1.23 & 0.03 & 1 & 5.06 & -- & poor \\
3 & 22.88 & 0.02 & 1.13 & 0.04 & 1.45 & 0.03 & 1 & 3.41 & -- & good \\
4 & 19.88 & 0.02 & 1.04 & 0.03 & 1.28 & 0.03 & 1.25 & 3.23 & -- & conf \\
5 & 22.78 & 0.02 & 0.81 & 0.03 & 1.14 & 0.03 & 1 & 2.9 & -- & good \\
6 & 22.66 & 0.02 & 1.04 & 0.03 & 1.17 & 0.03 & 1.25 & 6.3 & -- & good \\
7 & 22.17 & 0.02 & 0.88 & 0.03 & 1.16 & 0.03 & 1 & 6.71 & -- & good \\
8 & 21.73 & 0.02 & 1.18 & 0.03 & 1.48 & 0.03 & 1.5 & 16.82 & -- & fair \\
9 & 22.43 & 0.02 & 0.95 & 0.04 & 1.31 & 0.03 & 1 & 7.78 & -- & good \\
10 & 22.75 & 0.02 & 1.13 & 0.03 & 1.39 & 0.03 & 0.75 & 8.24 & -- & good \\
...\\
\enddata


\tablecomments{"Flag" indicates which bands are affected by the object being at the edge of a frame or between frames.  For the "Qual." tags, conf = spectroscopically confirmed GC; good = within 3 $\sigma$ color and size ranges defined using confirmed clusters; fair = within 3 $\sigma$ color ranges but not 3 $\sigma$ size range; poor = outside of 3 $\sigma$ color ranges; and nongc = spectroscopically confirmed as a non-GC.}


\end{deluxetable}

\clearpage

\begin{deluxetable}{cccccccc}

\tablenum{4}
\setlength{\tabcolsep}{0.05in}
\tablecaption{Sizes and Projected Galactocentric Distances of Globular Cluster Candidates}
\tablehead{\colhead{ID} & \colhead{Quality} & \colhead{V} & \colhead{Proj. Dist.} & \colhead{FWHM(major)} & \colhead{Axis ratio} & \colhead{$R{_h}$} & \colhead{$\sigma_{Rh}$} \\ 
\colhead{} & \colhead{} & \colhead{(mag)} & \colhead{(kpc)} & \colhead{(pix)} & \colhead{} & \colhead{(pc)} & \colhead{(pc)} } 

\startdata
1 & poor & 22.33 & 10.22 & 0.43 & 0.76 & 0.46 & 0.38 \\
2 & poor & 22.71 & 9.50 & 4.00 & 0.81 & 4.43 & 1.40 \\
3 & good & 22.88 & 10.69 & 1.98 & 0.86 & 2.25 & 0.79 \\
4 & conf & 19.88 & 9.47 & 2.07 & 0.94 & 2.46 & 0.98 \\
5 & good & 22.78 & 8.67 & 0.86 & 0.93 & 1.02 & 0.51 \\
6 & good & 22.66 & 7.50 & 4.96 & 0.85 & 5.61 & 1.67 \\
7 & good & 22.17 & 8.03 & 4.80 & 0.96 & 5.75 & 1.76 \\
8 & fair & 21.73 & 10.20 & 10.98 & 0.99 & 13.36 & 5.54 \\
9 & good & 22.43 & 10.13 & 6.67 & 0.85 & 7.55 & 3.09 \\
10 & good & 22.75 & 7.25 & 5.42 & 0.83 & 6.07 & 1.76 \\
...\\
\enddata
\end{deluxetable}
\clearpage

\end{document}